\documentclass[fleqn,usenatbib]{mnras}

\usepackage{newtxtext,newtxmath}
\usepackage[T1]{fontenc}

\DeclareRobustCommand{\VAN}[3]{#2}
\let\VANthebibliography\thebibliography
\def\thebibliography{\DeclareRobustCommand{\VAN}[3]{##3}\VANthebibliography}

\usepackage{graphicx}	% Including figure files
\usepackage{amsmath}	% Advanced maths commands
\usepackage{longtable} %DIF > 
\title[WALLABY Tully Fisher]{WALLABY Pre-Pilot and Pilot Survey: the Tully Fisher
Relation in Eridanus, Hydra, Norma and NGC4636 fields}

\author[H. Courtois et al.]
{\parbox{\textwidth}{H\'el\`ene M. Courtois$^{1}$\thanks{E-mail: helene.courtois@univ-lyon1.fr}, Khaled Said$^{2}$%\thanks{E-mail: k.saidahmedsoliman@uq.edu.au}
, Jeremy Mould$^{3}$%\thanks{E-mail: jmould@swin.edu.au}
, T.H. Jarrett$^4$, Daniel Pomar\`ede$^5$, Tobias Westmeier$^{6,7}$, Lister Staveley-Smith$^{6,7}$, Alexandra Dupuy$^{1,22}$, Tao Hong$^8$,  Daniel Guinet$^1$,  Cullan %DIF > 
Howlett$^2$,  
Nathan Deg$^{9}$, Bi-Qing For$^6$, Dane Kleiner$^{10}$, B\"arbel Koribalski$^{11,12}$, Karen Lee-Waddell$^{6,15}$, Jonghwan Rhee$^{6,7}$, Kristine Spekkens$^{13}$, Jing Wang$^{14}$, O.I. Wong$^{15,6,7}$, %DIF > 
Frank Bigiel$^{16}$, Albert Bosma$^{17}$, Matthew Colless$^{18}$, Tamara Davis$^2$, Benne Holwerda$^{19}$, Igor Karachentsev$^{20}$, Ren\'ee~C.~Kraan-Korteweg$^{4}$, Kristen B.W. McQuinn$^{21}$, Gerhardt Meurer$^{6}$, Danail Obreschkow$^{6,7}$, Edward Taylor$^3$} 
\vspace{0.4cm}\\
\parbox{\textwidth}{
$^{1}$University Lyon 1, IUF, IP2I Lyon, 69622 Villeurbanne cedex, France\\
$^{2}$School of Mathematics and Physics, The University of Queensland, Brisbane, QLD 4072, Australia\\
$^{3}$Centre for Astrophysics \& Supercomputing, Swinburne University, Australia\\
$^4$Astronomy Department,  University of Cape Town, Private Bag X3, Rondebosch, 7701, South Africa\\
$^5$IRFU, CEA Université Paris-Saclay, 91191 Gif-sur-Yvette, France\\
$^6$ International Centre for Radio Astronomy Research (ICRAR), The University of Western Australia, M468, Crawley, WA 6009, Australia\\
$^7$ARC Centre of Excellence for Astrophysics in 3 Dimensions (ASTRO 3D), Australia\\
$^8$National Astronomical Observatories, Chinese Academy of Sciences, 20A Datun Road, Chaoyang District, Beijing 100012, China\\ 
$^{9}$Department of Physics, Engineering Physics and Astronomy Queen's University Kingston, ON K7L 3N6, Canada\\ %DIF > 
$^{10}$INAF - Osservatorio Astronomico di Cagliari, Via della Scienza 5, 09047 Selargius, CA, Italy\\ %DIF > 
$^{11}$Australia Telescope National Facility, CSIRO Astronomy and Space Science, P.O. Box 76, NSW 1710, Epping, Australia\\ %DIF > 
$^{12}$Western Sydney University, Locked Bag 1797, Penrith, NSW 2751, Australia\\ %DIF > 
$^{13}$Royal Military College of Canada, PO Box 17000, Station Forces, Kingston, Ontario, Canada K7K7B4\\ %DIF > 
$^{14}$Kavli Institute for Astronomy and Astrophysics, Peking University, Beijing 100871, China\\ %DIF > 
$^{15}$CSIRO Space \& Astronomy, PO Box 1130, Bentley, WA 6102, Australia\\ %DIF > 
$^{16}$Argelander-Institut f\"ur Astronomie, Universit\"at Bonn, Auf dem H\"ugel 71, 53121 Bonn, Germany\\ %DIF > 
$^{17}$Aix Marseille Univ, CNRS, CNES, LAM, Marseille, France\\ %DIF > 
$^{18}$Research School of Astronomy and Astrophysics, Australian National University, Canberra, ACT 2611, Australia\\ %DIF > 
$^{19}$University of Louisville, Department of Physics and Astronomy, 102 Natural Science Building, 40292 KY Louisville, USA\\  
$^{20}$Special Astrophysical Observatory, Russian Academy of Sciences, Russia\\ %DIF > 
$^{21}$Rutgers University, Department of Physics and Astronomy, 136 Frelinghuysen Road, Piscataway, NJ 08854, USA\\ %DIF > 
$^{22}$Korea Institute for Advanced Study, 85, Hoegi-ro, Dongdaemun-gu, Seoul 02455, Republic of Korea   }}

\date{Accepted October 19th 2022 for MNRAS}

\pubyear{2022}

\RequirePackage[normalem]{ulem} 
\RequirePackage{color}\definecolor{RED}{rgb}{1,0,0}\definecolor{BLUE}{rgb}{0,0,1}

\providecommand{\DIFaddbegin}{} %DIF PREAMBLE
\providecommand{\DIFaddend}{} %DIF PREAMBLE
\providecommand{\DIFdelbegin}{} %DIF PREAMBLE
\providecommand{\DIFdelend}{} %DIF PREAMBLE
\providecommand{\DIFaddbeginFL}{} %DIF PREAMBLE
\providecommand{\DIFaddendFL}{} %DIF PREAMBLE
\providecommand{\DIFdelbeginFL}{} %DIF PREAMBLE
\providecommand{\DIFdelendFL}{} %DIF PREAMBLE
\newcommand{\DIFscaledelfig}{0.5}
\RequirePackage{settobox} %DIF PREAMBLE
\RequirePackage{letltxmacro} %DIF PREAMBLE
\newsavebox{\DIFdelgraphicsbox} %DIF PREAMBLE
\newlength{\DIFdelgraphicswidth} %DIF PREAMBLE
\newlength{\DIFdelgraphicsheight} %DIF PREAMBLE
\LetLtxMacro{\DIFOincludegraphics}{\includegraphics} %DIF PREAMBLE
\newcommand{\DIFaddincludegraphics}[2][]{{\color{blue}\fbox{\DIFOincludegraphics[#1]{#2}}}} %DIF PREAMBLE
\newcommand{\DIFdelincludegraphics}[2][]{% %DIF PREAMBLE
\sbox{\DIFdelgraphicsbox}{\DIFOincludegraphics[#1]{#2}}% %DIF PREAMBLE
\settoboxwidth{\DIFdelgraphicswidth}{\DIFdelgraphicsbox} %DIF PREAMBLE
\settoboxtotalheight{\DIFdelgraphicsheight}{\DIFdelgraphicsbox} %DIF PREAMBLE
\scalebox{\DIFscaledelfig}{% %DIF PREAMBLE
\parbox[b]{\DIFdelgraphicswidth}{\usebox{\DIFdelgraphicsbox}\\[-\baselineskip] \rule{\DIFdelgraphicswidth}{0em}}\llap{\resizebox{\DIFdelgraphicswidth}{\DIFdelgraphicsheight}{% %DIF PREAMBLE
\setlength{\unitlength}{\DIFdelgraphicswidth}% %DIF PREAMBLE
\begin{picture}(1,1)% %DIF PREAMBLE
\thicklines\linethickness{2pt} %DIF PREAMBLE
{\color[rgb]{1,0,0}\put(0,0){\framebox(1,1){}}}% %DIF PREAMBLE
{\color[rgb]{1,0,0}\put(0,0){\line( 1,1){1}}}% %DIF PREAMBLE
{\color[rgb]{1,0,0}\put(0,1){\line(1,-1){1}}}% %DIF PREAMBLE
\end{picture}% %DIF PREAMBLE
}\hspace*{3pt}}} %DIF PREAMBLE
} %DIF PREAMBLE
\LetLtxMacro{\DIFOaddbegin}{\DIFaddbegin} %DIF PREAMBLE
\LetLtxMacro{\DIFOaddend}{\DIFaddend} %DIF PREAMBLE
\LetLtxMacro{\DIFOdelbegin}{\DIFdelbegin} %DIF PREAMBLE
\LetLtxMacro{\DIFOdelend}{\DIFdelend} %DIF PREAMBLE
\DeclareRobustCommand{\DIFaddbegin}{\DIFOaddbegin \let\includegraphics\DIFaddincludegraphics} %DIF PREAMBLE
\DeclareRobustCommand{\DIFaddend}{\DIFOaddend \let\includegraphics\DIFOincludegraphics}

\DeclareUnicodeCharacter{2212}{-}%DIF PREAMBLE
\DeclareRobustCommand{\DIFdelbegin}{\DIFOdelbegin \let\includegraphics\DIFdelincludegraphics} %DIF PREAMBLE
\DeclareRobustCommand{\DIFdelend}{\DIFOaddend \let\includegraphics\DIFOincludegraphics} %DIF PREAMBLE
\LetLtxMacro{\DIFOaddbeginFL}{\DIFaddbeginFL} %DIF PREAMBLE
\LetLtxMacro{\DIFOaddendFL}{\DIFaddendFL} %DIF PREAMBLE
\LetLtxMacro{\DIFOdelbeginFL}{\DIFdelbeginFL} %DIF PREAMBLE
\LetLtxMacro{\DIFOdelendFL}{\DIFdelendFL} %DIF PREAMBLE
\DeclareRobustCommand{\DIFaddbeginFL}{\DIFOaddbeginFL \let\includegraphics\DIFaddincludegraphics} %DIF PREAMBLE
\DeclareRobustCommand{\DIFaddendFL}{\DIFOaddendFL \let\includegraphics\DIFOincludegraphics} %DIF PREAMBLE
\DeclareRobustCommand{\DIFdelbeginFL}{\DIFOdelbeginFL \let\includegraphics\DIFdelincludegraphics} %DIF PREAMBLE
\DeclareRobustCommand{\DIFdelendFL}{\DIFOaddendFL \let\includegraphics\DIFOincludegraphics} %DIF PREAMBLE
\begin{document}
\label{firstpage}
\pagerange{\pageref{firstpage}--\pageref{lastpage}}

\maketitle

\vspace{-4cm}
\begin{abstract}
The WALLABY pilot survey has  been conducted using the Australian SKA Pathfinder (ASKAP).
The integrated 21-cm HI line spectra are formed in a very different manner compared to usual single-dish spectra Tully-Fisher measurements. It is thus extremely important to ensure that slight differences (e.g. biases due to missing flux) are quantified and understood in order to maximise the use of the large amount of data becoming available soon.
This article is based on four fields for which the data are scientifically interesting by themselves.
The pilot data discussed here consist of 614 galaxy spectra at a rest wavelength of 21cm. Of these spectra, 472 are of high enough quality to be used to potentially derive distances using the Tully-Fisher relation. 
We further restrict the sample to the 251 galaxies whose inclination is sufficiently close to edge-on. For these, we derive Tully-Fisher distances using the deprojected WALLABY velocity widths combined with infrared (WISE W1) magnitudes. The resulting Tully-Fisher distances for the Eridanus, Hydra, Norma and NGC 4636 clusters 
are 21.5, 53.5, 69.4 and 23.0 Mpc respectively, with uncertainties of 5--10\%, which are better or equivalent to the ones obtained in studies using data obtained with giant single dish telescopes.
The pilot survey data show the benefits of WALLABY over previous giant single-dish telescope  surveys.
WALLABY is expected to detect around half a million galaxies with a mean redshift of $z = 0.05 (200 Mpc)$.  This study suggests that about 200,000 Tully-Fisher distances might result from the survey.

\end{abstract}


%%%%%%%%%%%%%%%%%%%%%%%%%%%%%%%%%%%%%%%%%%%%%%%%%%

%%%%%%%%%%%%%%%%% BODY OF PAPER %%%%%%%%%%%%%%%%%%
%\clearpage
\section{INTRODUCTION}
\label{sec:intro}

\cite{Tully:1977} discovered a relation (the Tully-Fisher relation: TFR) between the extensive properties of disk galaxies (mass, luminosity, and radius) and their rotation velocity, which can be derived from the velocity width of their neutral hydrogen 21 cm spectra.
If galaxy surface brightness, stellar mass, and mass to light ratio are
functions of halo mass, the virial theorem leads one to expect a luminosity -- 
rotation velocity relation. As the TFR relates distance-independent and distance-dependent quantities, it can be used to measure galaxy distances
and peculiar velocities. Developments in understanding 
galaxy scaling relations are reviewed by \cite{Mould:2020}. The TFR is also reproduced in dark matter simulations, when sub-grid physics is added to follow the baryons (e.g. \citealt{Springel:2018}), so it can therefore also be used to further probe the astrophysics of galaxy formation. 

The Widefield ASKAP L-band Legacy All-sky Blind Survey (WALLABY) pre-pilot and pilot phase 1 survey is an observational study of four fields with the Australian Square Kilometer Array Pathfinder (ASKAP) telescope. The pilot observations are a precursor to the full WALLABY survey, a blind neutral hydrogen survey of the
Southern hemisphere \citep{Koribalski:2020} which is scheduled to begin in 2022. The pilot data has already been used to study HI content of galaxies in groups \citep{2021MNRAS.507.2300F}, discover dark clouds in the vicinity of galaxies \citep{2021MNRAS.507.2905W}, and study the diversity of ram pressure stripping of HI gas in galaxies in the Hydra cluster \citep{2021ApJ...915...70W}.

 The raw ASKAP visibilities from the WALLABY observations are processed and reduced using ASKAPsoft. The processed  data products are available on CASDA. The data are first released to the WALLABY team for analysis, and, once published, available to the wider community. A mosaic of individual footprints is then made to produce the final full sensitivity cubes which are run through the SoFiA pipeline. SoFiA is a 3D source finding and parametrisation software package \citep{Serra:2015, Westmeier2021}   which, amongst many other outputs, produces a 1-dimensional spectrum by spatially integrating all emission within its mask. The measured HI linewidth is a measure of the galaxy projected rotation velocity. 

In this article we explore the infrared TFR, by combining WALLABY pilot HI spectra with photometric data using WISE W1 band infrared fluxes.
Infrared wavelengths have the advantage of markedly lower extinction of galaxy radiation by dust \citep{Aaronson:1979}. It is now well known that using infrared data provides  a tighter TFR than using optical bands \citep{2013ApJ...765...94S, 2014ApJ...792..129N, 2020ApJ...902..145K}.
The measured galaxy distances can be used to: measure the expansion rate through determination of the Hubble constant \citep{2020ApJ...902..145K}; test the $\Lambda$CDM cosmological model using measurements of the bulk flow \citep{2015MNRAS.449.4494H, 2019MNRAS.487.2061H}; compute systematics in the SNIa Hubble constant determination \citep{2021arXiv211003487P}. Another question that measurements of non-Hubble flows can address is the quantity of matter needed in specific regions such as the Great Attractor in order to explain gravitationally-induced motions, with the aim of reconciling with observed baryonic matter distribution \citep{Said2020,2020RNAAS...4..159T}.
%, only global dynamics cosmography computation can reconcile dark and baryonic matter distributions.

A major benefit of the WALLABY survey is its higher sensitivity compared to previous southern surveys used for Tully-Fisher distances -- for example, the HI Parkes All-Sky Survey (HIPASS) \citep{2004AJ....128...16K, 2004MNRAS.350.1195M}. It also has higher survey speed than northern single-dish telescopes previously used to conduct HI surveys \citep{2018ApJ...861...49H} thanks to its wide (30 sq.deg) instantaneous field of view, thus enabling to construct larger galaxy samples. Other advantages are the location of ASKAP at a site that is free from terrestrial radio frequency interference and its high angular resolution, which allows detailed morphological and dynamical studies. Although angular resolution is not important for this TFR study, it does lessen the probability of beam confusion and will allow future dynamical measurements of inclination.
%, enabled by the phased array feeds. Thus, in relation to TFR studies, the fact that it can observe and detect so many galaxies in so few observations and observing time is its greatest strength.   

All-sky infrared surveys provide complementary data for large samples of galaxies with a depth, resolution and photometric accuracy suitable for TFR studies. Examples include the 2MASS Tully-Fisher (2MTF) survey of 2,062 galaxies \citep{2019MNRAS.487.2061H} and the WISE Tully-Fisher study of \citet{2013ApJ...771...88L}. Currently, the largest Tully-Fisher samples are of the order of 10,000 galaxies \citep{2020ApJ...902..145K}, while WALLABY is planned to triple or more this latter number in the next few years \citep{Koribalski:2020}. The Meerkat SKA precursor located in South Africa is also used to successfully derive baryonic Tully-Fisher galaxy distances \citep{2021MNRAS.508.1195P}.

Galaxy distances are key to the objective of the CosmicFlows (CF) project \citep{Tully2016}, which is to improve the cosmography of the nearby universe including the location and mass measurement 
%by paying special attention to locating and measuring the full mass 
of the largest attractors. Distances are used to compute velocity deviations from the Hubble cosmic expansion which are assumed to arise from the large-scale gravitational field. This can only be realized with very accurate distances. The CF project started in 2006 and will shortly publish its fourth generation catalogue containing about 45,000 accurate determinations of galaxy distances.

In this paper we describe the pre-pilot and phase 1 pilot WALLABY surveys from which our HI data is drawn, and present optical identifications of the detected sourced at 21 cm ($\S2)$; we form TFRs by using accurate photometry in the WISE W1 bandpass ($\S3$). Finally, we show how the WALLABY Survey can add to the CF database for future studies of large-scale structure.

\section{WALLABY pre-pilot and pilot survey}

\begin{table}
\caption{Table of J2000 coordinates pointings of the WALLABY pre-pilot and pilot phase 1 survey fields.}
\begin{center}
\begin{tabular}{lcc}
\hline
Footprint&\multicolumn{2}{c}{Field centre (J2000)}\\
     &     RA    &    DEC\\
\hline
Eridanus A & 03:39:30.000 & -22:23:00.00 \\
Eridanus B & 03:36:44.520 & -22:37:54.69 \\
%Eridanus super group & 03:38:00 &-22:30:00 \\
Hydra-1A &	10:15:47.844 & -27:22:27.66 \\
Hydra-1B &	10:17:49.958 & -27:49:24.30\\
Hydra-2A &	10:39:24.238 & -27:22:27.66\\
Hydra-2B &	10:41:26.352 &-27:49:24.30\\
NGC4636-1A &	12:38:02.328 &-00:26:59.95\\
NGC4636-1B &	12:39:50.337 &-00:53:59.85\\
NGC4636-2A &	12:38:02.729 &+04:56:58.64\\
NGC4636-2B &	12:39:51.059 &+04:29:58.13\\
Norma-1A &	16:16:30.928 &-59:27:41.88\\
Norma-1B &	16:20:06.332 &-59:54:30.90\\
Norma-2A &	16:55:26.063 &-59:27:41.88\\
Norma-2B &	16:59:01.467 &-59:54:30.90\\
\hline
\end{tabular}
\label{centers}
\end{center}
\label{table_footprints}
\end{table}
%\textcolor{red}{Karen Lee-Waddel: Table 1 : You might want to include the SB (scheduling block) IDs  so a reader could easily find the data on CASDA. Note: the released data does have a DOI (one for pre-pilot and one for Phase 1), it would be appropriate to include those in the paper, perhaps as a footnote.}	

WALLABY is an extragalactic all-sky HI survey with the ASKAP telescope which consists of 36 × 12 m dishes with baselines ranging from 22 m to 6 km \citep{Hotan2021}. It is equipped with phased array feeds with a field-of-view of about 30 deg² at 1.4 GHz. The frequency coverage of ASKAP is from 0.7 to 1.8 GHz with an instantaneous bandwidth of 288 MHz. Its spectral resolution of 18.5 kHz translates into a velocity resolution of 4 km~s$^{-1}$ at redshift zero.

For the WALLABY pilot surveys, a bandwidth of 144 MHz covering the frequency range of 1295.5 to 1439.5 MHz, was processed and imaged. This corresponds to redshifts $z < 0.096$. All baselines, including the 6 km baselines, were calibrated. However, the current WALLABY images only include baselines up to 2 km, resulting in an angular resolution of 30\arcsec. The nominal RMS noise level is approximately 1.6 mJy after 2 footprints, each of duration 8 h, are combined. We refer the reader to \cite{Koribalski:2020} for more details about the WALLABY survey.

\subsection{Radio and photometry data measurements}

The pre-pilot and pilot phase 1 survey resulted in coverage of the NGC4636 group, the Eridanus super-group, the Hydra I cluster, and the Norma cluster surrounds (the actual Norma cluster is excluded from this study due to strong residual radio continuum emission). Sources were catalogued using the SoFIA source finding pipeline \citep{Serra:2015,Westmeier2021}.
Detections were linked across a spatial and spectral radius of 2 pixels with a minimum size requirement for reliable source identification of 5 spatial pixels and 8 spectral channels. SoFiA’s reliability filter was then applied to remove all detections with a reliability below 0.8, using a Gaussian kernel density estimator of size 0.35 times the covariance. All remaining sources were then parameterised, assuming a restoring beam size of 30\arcsec\ for all integrated flux measurements. The source detection details slightly vary between the four different fields -- details will be provided in the upcoming paper accompanying the first public data release of phase 1 pilot data (Westmeier et al. in prep. 2022).

The pre-pilot field containing the Eridanus super group was observed with two interleaving footprints (A and B) with an offset of 38\arcmin, and rotated by a position angle of 45 deg relative to a standard
ASKAP footprint. The Norma, Hydra and NGC4636 pilot observations were of two adjacent 30 sq.deg overlapping tiles (1 and 2) with the tile centres separated by 5.4 deg. Each tile consisted of two slightly offset footprints observed for 8 hours. Therefore, each 1-D HI spectrum has an effective integration time of 16 hr. The centres of the footprints are given in Table \ref{centers}.

\begin{figure}
\begin{center}
\includegraphics[width=1\columnwidth,angle=-0]{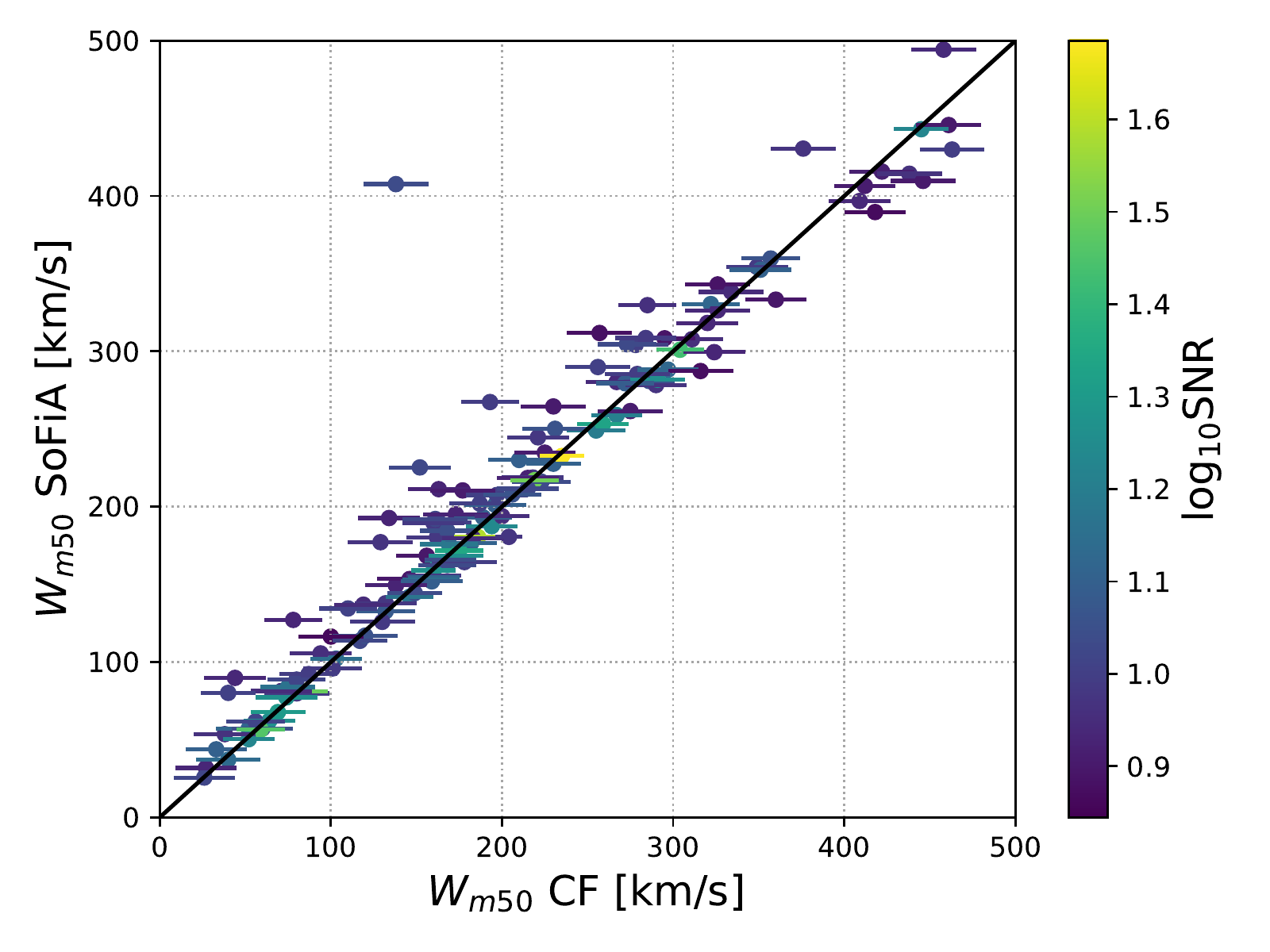}
\caption{Linewidths W$_\mathrm{m50}$ in the Norma field as measured by the CosmicFlows (CF) algorithm compared with the automatic measurements of SoFiA, colour-coded by S/N ratio. The plot only includes spectra with an error on W$_\mathrm{m50}$ less or equal to 20~km s$^{-1}$. 
%Few discrepant data are found with the SoFiA release. 
Two further spectra with low S/N ratio, where the linewidth difference exceeds 100 km s$^{-1}$, are excluded from the plot. }
\label{fig:sofia}
\end{center}
\end{figure}

\subsection{Comparison of velocity width measurement algorithms}

\begin{table*}
\begin{center}
\caption[]{All galaxies with pairwise relative error $|\epsilon| > 1.0$ between the CosmicFlows and S16 algorithms.}
\begin{tabular} {lcccccc}
\hline\hline
name & PGC & $W_{m50}$ & $e_{Wm50}$ & $W_{m50}^{S16}$ & $e_{Wm50}^{S16}$ & $\epsilon$\\
\hline
J102439-244547 & 30532 &  94 &  18 & 112 &   2 &  $-1.0177$\\
J102447-264054 & 30545 & 271 &  17 & 322 &  13 &  $-2.4078$\\
J103442-283406 & 31296 & 247 &  15 & 263 &   4 &  $-1.0149$\\
J103726-261843 & 31557 & 199 &  18 & 224 &   2 &  $-1.3854$\\
J104127-275119 & 31842 &  99 &  19 & 119 &   5 &  $-1.0091$\\
J104142-284653 & 31855 & 107 &  15 & 128 &   7 &  $-1.2566$\\
J104513-262755 &  &  50 &  17 &  74 &   6 &  $-1.3571$\\
J104905-292232 & 32361 & 117 &  17 & 142 &   3 &  $-1.4259$\\
J163452-603705 & 58536 & 189 &  17 & 215 &   1 &  $-1.5087$\\
J164359-600237 &  & 215 &  18 & 244 &  17 &  $-1.1702$\\
J165105-585918 & 59112 &  84 &  14 & 114 &   3 &  $-2.0721$\\
J165145-590915 &  &  71 &  18 &  90 &   5 &  $-1.0111$\\
\hline
\end{tabular}
\label{large_discrepancies_CF_vs_S16}
\end{center}
\end{table*}

SoFiA provides approximate linewidths for all galaxies that it finds in WALLABY cubes. However, to improve robustness, accuracy, reliability and compatibility with previous measurements, we also compute linewidths using the CF methodology and make a detailed comparison with the SoFiA estimates.
 %Following methods adopted in the CosmicFlows program, we measure the neutral hydrogen linewidth at 21cm wavelength enclosing 50\% of the cumulative HI line flux, $W_\mathrm{m50}$ \citep{Courtois:2011b,Courtois:2011a}.
 The CF linewidth $W_\mathrm{m50}$, is measured at the flux level that is 50\% of the mean flux, averaged in channels within the wavelength range enclosing 90\% of the total integrated flux \citep{Courtois:2011b,Courtois:2011a}. %However, the parameter $W_\mathrm{m50}$ is only an empirical measure of the true width of an HI galaxy velocity profile. 
 It is followed by a correction for redshift and instrumental broadening: $W^\mathrm{corr}_\mathrm{m50} = (W_\mathrm{m50}- 2\Delta v \lambda)/(1 +z) $, where $z$ is the redshift in the CMB frame, $\Delta v$ is the smoothed spectral resolution, and $\lambda = 0.25$ is an empirically determined constant \citep{Courtois:2009}. The observed line width was also adjusted by separating out the broadening from turbulent motions and offsets to produce an approximation to $2 V_\mathrm{max}$, where $V_\mathrm{max}$
 = W$_{mx}$/2 correspond to the rotation rate over the main body of a galaxy. \citet{Tully:1985} define the parameter $W_\mathrm{mx}$ as:

\begin{equation}
\begin{aligned}
    W^2_\mathrm{mx}= & W^\mathrm{corr,2}_\mathrm{m50} + W^2_\mathrm{t,m50} \left[ 1 - 2 e^{-(W^\mathrm{corr}_\mathrm{m50}/W_\mathrm{c,m50})^2} \right] \\
    & - 2 W^\mathrm{corr}_\mathrm{m50} W_\mathrm{t,m50} \left[ 1 - e^{-(W^\mathrm{corr}_\mathrm{m50}/W_\mathrm{c,m50})^2} \right].
\end{aligned}
\label{eq:wmx}
\end{equation}

The parameters $W_\mathrm{c,m50}= 100$ km s$^{-1}$ and $W_\mathrm{t,m50}= 9$ km s$^{-1}$ are set following tests conducted in \cite{Courtois:2009}, and characterize respectively the transition from boxcar to Gaussian intrinsic profiles, and the turbulent broadening for the observed linewidth considered. It is then related to the rotation rate $V_\mathrm{max}$ by:
\begin{equation}
    V_\mathrm{max}= \frac{W_\mathrm{mx}}{2 \sin{(i)}},
\end{equation}
where $i$ is the inclination of the galaxy from face-on relative to the observer.

Details regarding the $W_\mathrm{m50}$ and $W_\mathrm{mx}$ line width parameters and comparisons with alternatives are discussed in \cite{Courtois:2009}. 
%For our further TFR calculations, these values replace the linewidths supplied by the SoFiA pipeline in the pilot survey data release (see Figure \ref{fig:sofia}).
A comparison of the CF and SoFiA linewidth measurements in the Norma field is shown in Fig.~\ref{fig:sofia}.

\begin{figure}
\includegraphics[width=\columnwidth,angle=-0]{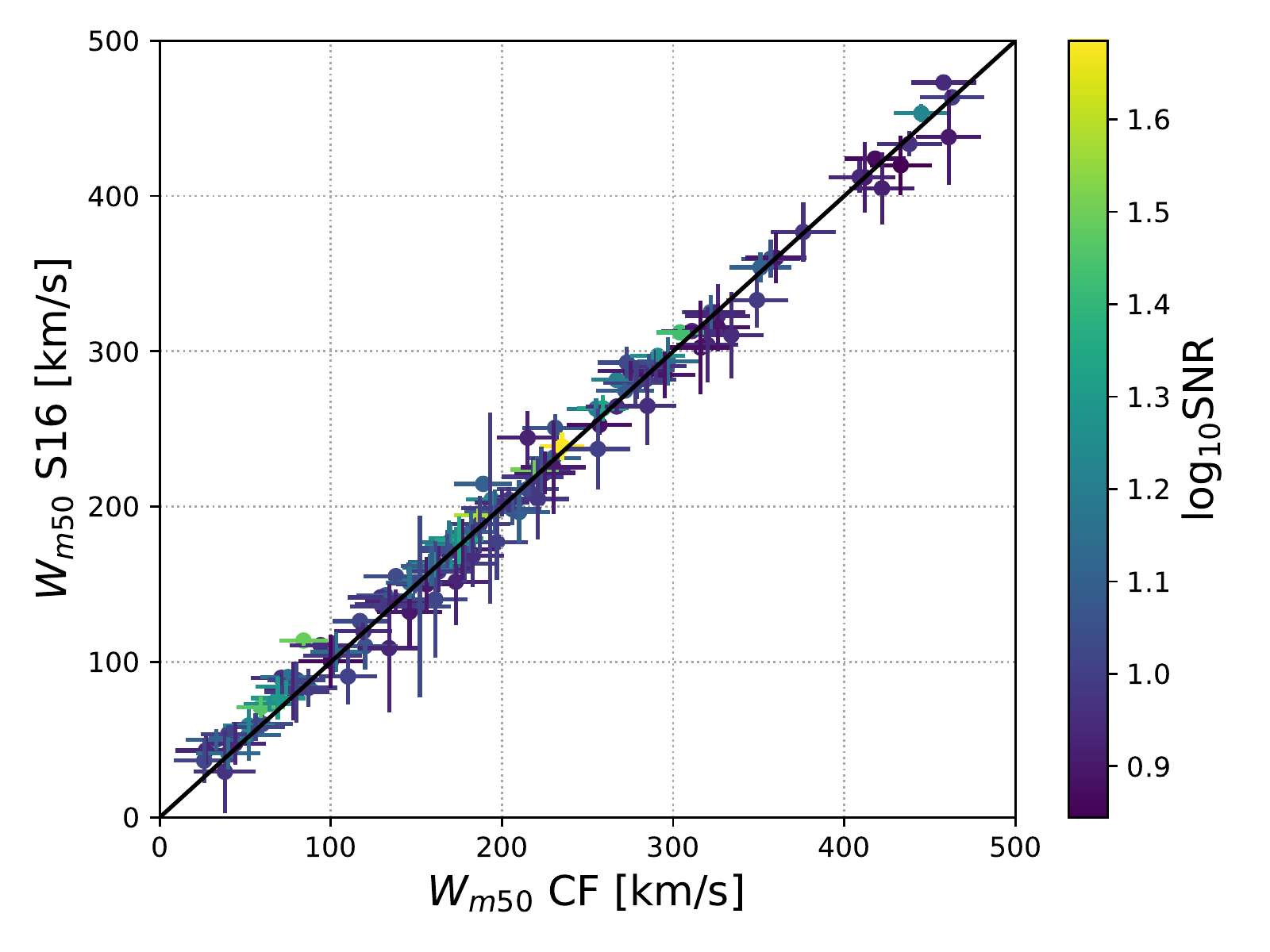}
\caption{Comparison of 50\% linewidths in the Norma field measured with the CF algorithm and linewidths measured at 50\% of the mean flux using 
the S16 algorithm \protect\citep{Said2016}. Only measurements of spectra with an adequate quality are displayed (error on W$_\mathrm{m50}$ less or equal to 20 km s$^{-1}$). This comparison displays a typical scatter at 20 kms $^{-1}$.}
\label{fig:khaled}
\end{figure}

We make a further comparison with velocity widths derived using the method described by \cite{Said2016} (hereafter S16). This algorithm requires prior knowledge of the left and right horns which is currently done manually and returns the linewidth measured at 50\% of the mean flux and its associated uncertainty. In this paper, we compare the CF linewidth with the S16 linewidth measured at 50\% of the mean flux, as shown in Fig.~\ref{fig:khaled}.

%We adopt that used by the CosmicFlows collaboration, which has been corrected following equation (1).

We use the relative error between pairs of measurements to quantitatively check for any large discrepancies between 50\% widths measured with the CF algorithm and widths measured at 50\% of the mean flux with the S16 algorithm. We used the pairwise relative error of the form:
\begin{equation}
    \varepsilon = \frac{W_\mathrm{m50}^\mathrm{CF}-W_\mathrm{m50}^{\mathrm{S16} } }{ \sqrt{ ({eW_\mathrm{m50}^\mathrm{CF}}^2 + {eW_\mathrm{m50}^\mathrm{S16}}^2) }},
\end{equation}
where $W_\mathrm{m50}^\mathrm{CF}$, $W_\mathrm{m50}^\mathrm{S16}$, $eW_\mathrm{m50}^\mathrm{CF}$, and $eW_\mathrm{m50}^\mathrm{S16}$ are the 50\% widths measurements from CosmicFlows and S16 algorithms and their associated statistical uncertainties, respectively. Values of $\varepsilon$ for all galaxies are within the distribution of a Gaussian with a mean of zero and a standard deviation of unity. Table \ref{large_discrepancies_CF_vs_S16} lists all discrepant measurements with $|\varepsilon| > 1.0$.

Following the CosmicFlows collaboration, the signal-to-noise ratio is calculated by taking the mean flux density across all channels in excess of 95\% of the peak and dividing it by the noise level in the spectrum as measured outside of the HI line \citep{Courtois:2015}.
%It is not trivial to transfer this definition onto the 3D WALLABY spectra projected onto 1 dimension spectra since the mask size changes with frequency (and forces the noise to be zero beyond the mask). To be able to work with 1D SoFiA spectra we set the signal to noise value equal to the maximum flux level in the HI line.
%Also it could be questioned if this relation is true for measurements from all instruments, or if one should calibrate ASKAP separately (e.g. by external comparisons to line width measurements for the same galaxies). This is why we apply this relation and search for inconsistency with previously used instruments on the same galaxies (see the discussion with Figure~\ref{fig:compwmx}).\\

The error on the linewidth $e_W$ in km s$^{-1}$, follows the empirical relation from \citep{Courtois:2009}:
\begin{equation}
\begin{aligned}
    \mathrm{SNR} \geq 17 \;\;\;\; & e_W = 8 ~km\ s^{-1} ; \\
    2 < \mathrm{SNR} < 17 \;\;\;\; & e_W = 21.6 - 0.8 \times \mathrm{SNR} ~km\ s^{-1}; \\
    \mathrm{SNR} \leq 2 \;\;\;\; & e_W = 70 - 25 \times \mathrm{SNR} ~km\ s^{-1}, \\
\end{aligned}
\end{equation}
where SNR is the signal-to-noise ratio. CosmicFlows has shown that an HI spectrum is considered adequate for estimating galaxy distance through the TFR if the error on $W_\mathrm{m50}$ is less or equal to 20 km s$^{-1}$. As seen in Fig.~\ref{fig:khaled} this threshold is confirmed once again by this study.

In summary for the three algorithms for $W_\mathrm{m50}$: (1) SoFiA is automated but sometimes gives discrepant measurements for low SNR spectra (see Fig. 1). It also has a dispersion of about 50 km s$^{-1}$, which is too high for Tully-Fisher purposes \citep{Courtois:2009}. 
(2) The adopted CF algorithm is semi-automated, with a final inspection by eye of the fit. In some cases (six spectra, corresponding to about 1\% of all spectra in the pilot survey), a fit was not possible.
 %.  This happened in 1\% of the cases of  the pilot survey: 6 spectra.
(3) The S16 algorithm requires a starting value for $W_\mathrm{m50}$ (that can be provided by SoFiA), but appears to provide a measurement which is within the 20 km s$^{-1}$ error limit on $W_\mathrm{m50}$ that we impose for TFR studies. 
Therefore the WALLABY survey data will be treated by all three algorithms. The comparison of the results per individual galaxy will allow to identify the difficult spectra that require close inspection.

%This comparison of the three algorithms for the measurement of $W_\mathrm{m50}$ as seen in Figures \ref{fig:sofia} and \ref{fig:khaled}, shows  how robust the three algorithms are. They agree within the error threshold we impose for deriving Tully-Fisher distances for galaxies of 20 km s$^{-1}$. In the rest of the article, we will be using the $W_\mathrm{m50}$ values of CosmicFlows algorithm apart from a few galaxies that were best fitted by the S16 algorithm.

Details of all the galaxies selected at 21 cm for this TF study are listed in Appendix \ref{data}.

\subsection{Comparison with previous single-dish velocity width measurements}

In this section we check whether there are any differences between single-dish and multi-dish spectroscopic data, that would affect the quality of the derived Tully-Fisher distances with systematic errors.
%For example, missing flux in interferometer data may reduce the amount of extended flat-rotation curve gas in the reconstructed profile and decrease the velocity width measurement.

%Similarly, due to large single-dish beamwidths, HI detections may have wrong optical cross-identifications, typically wrongly identifying a host that is brighter and of larger mass than the correct identification.
%Galaxy cross-identification is also an external test of the robustness of matching optical counterparts to WALLABY spectra. 

\begin{figure}
\begin{center}
\includegraphics[width=1\columnwidth,angle=-0]{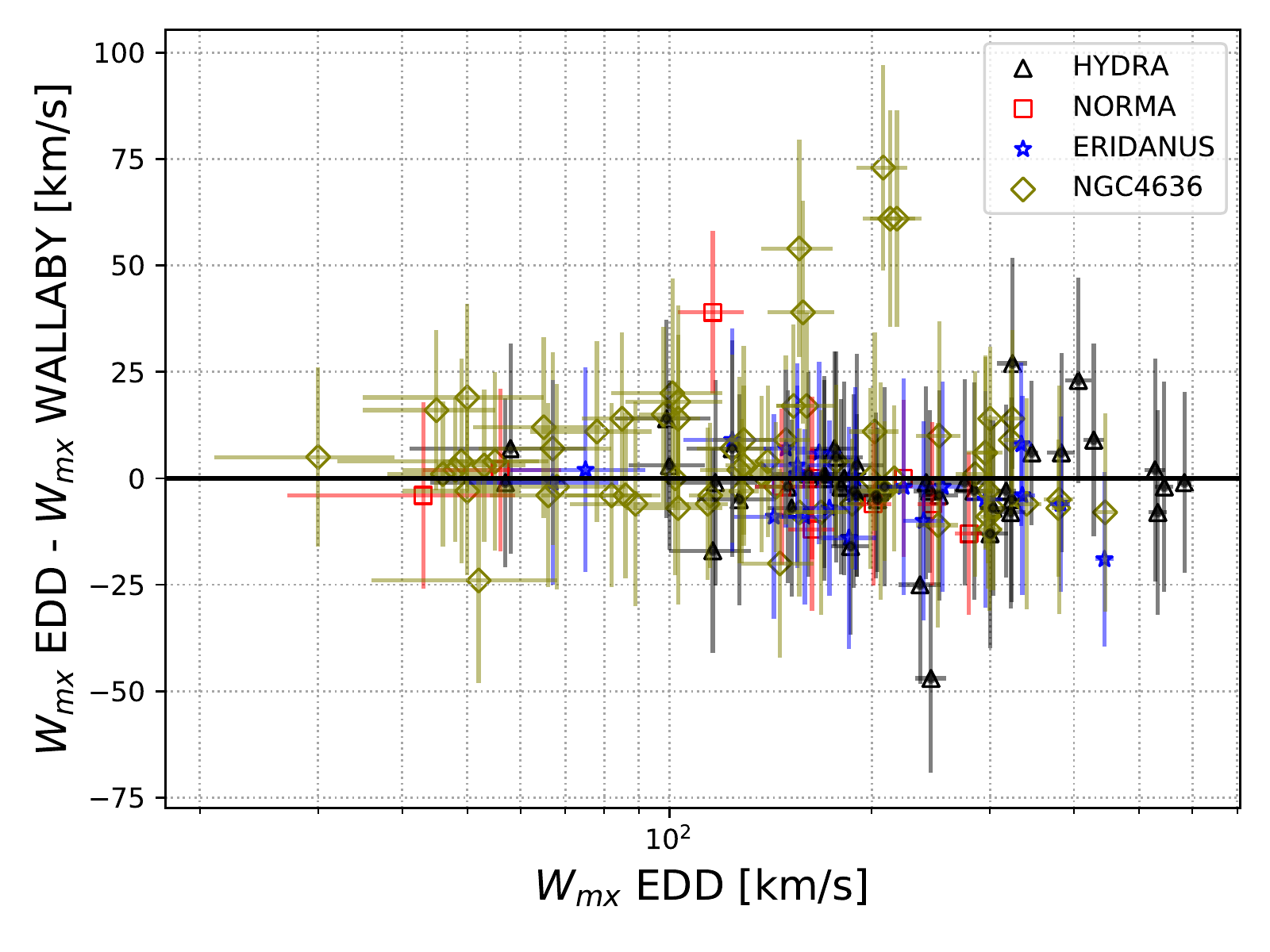}
\caption{Comparison of the WALLABY pilot phase velocity widths presented in this article with single-dish data in the  Extragalactic Distance
Database \citep{Courtois:2015}. Widths are only compared where the internal error for $W_\mathrm{m50}$ in both datasets is less or equal to 20 km s$^{-1}$. Measurements mostly agree within the 20~km s$^{-1}$ limit. The large offsets are either due to low SNR Arecibo spectra or due to noisy WALLABY spectra with residual baseline ripples due to residuals caused by strong radio continuum sources. The latter is  more prominent in the NGC4636 field, which lies close to 3C273.}
\label{fig:compwmx}
\end{center}
\end{figure}

Fortunately, there is a sufficient overlap with existing data to check consistency between pilot WALLABY and published single-dish velocity widths as compiled in the Extragalactic Distance Database \citep{Courtois:2015}. This comparison is shown in Fig.~\ref{fig:compwmx} for the 
galaxies in common. There is no significant difference when comparing the WALLABY W$_\mathrm{mx}$ values with those measured from single-dish data, with most agreeing within the 20~km s$^{-1}$ limit. The discrepant data comes from  wrong width measurements from  low S/N ratio Arecibo spectra and measurements of WALLABY spectra containing residual continuum from strong nearby sources outside the field and therefore not cleaned. This comparison is an excellent external robustness assessment of the capability of WALLABY to deliver satisfactory HI data to the community. 
It also demonstrates that the exposure time used for the Phase 1 Pilot Survey can be considered as the guideline for the upcoming large WALLABY survey.

\subsection{WISE photometry}\label{Sect: WISE Photometry}
The \textit{Wide-Field Survey Explorer} (WISE) satellite provides all-sky photometric observations in 4 mid-infrared bands, W1 = $3.4 \mu$m, W2 = $4.6 \mu$m, W3 = $12.0 \mu$m, W4 = $22.8 \mu$m \citep{Wright:2010}.

\begin{figure}
%\begin{center}
\includegraphics[width=\columnwidth]{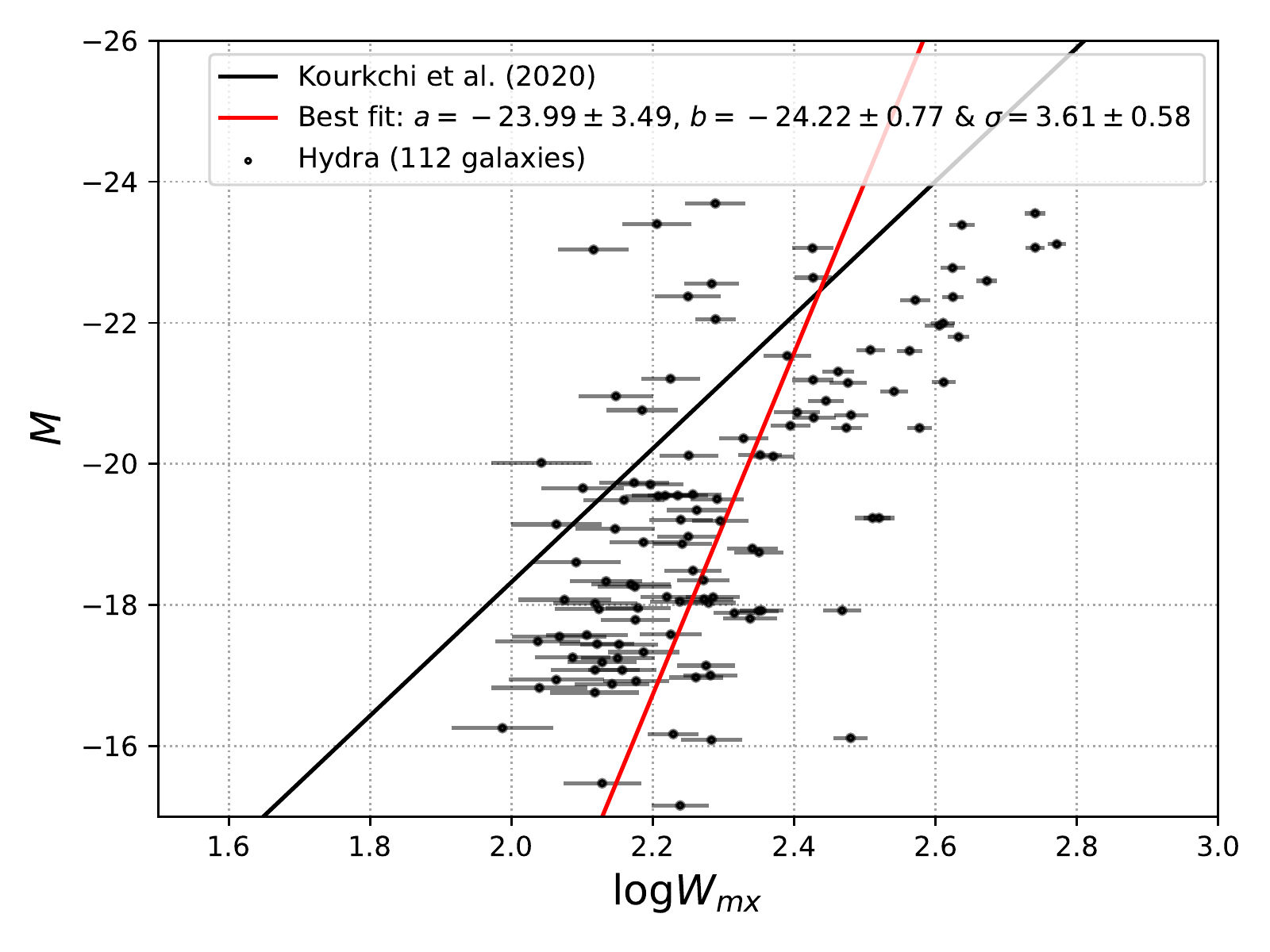}
\includegraphics[width=\columnwidth]{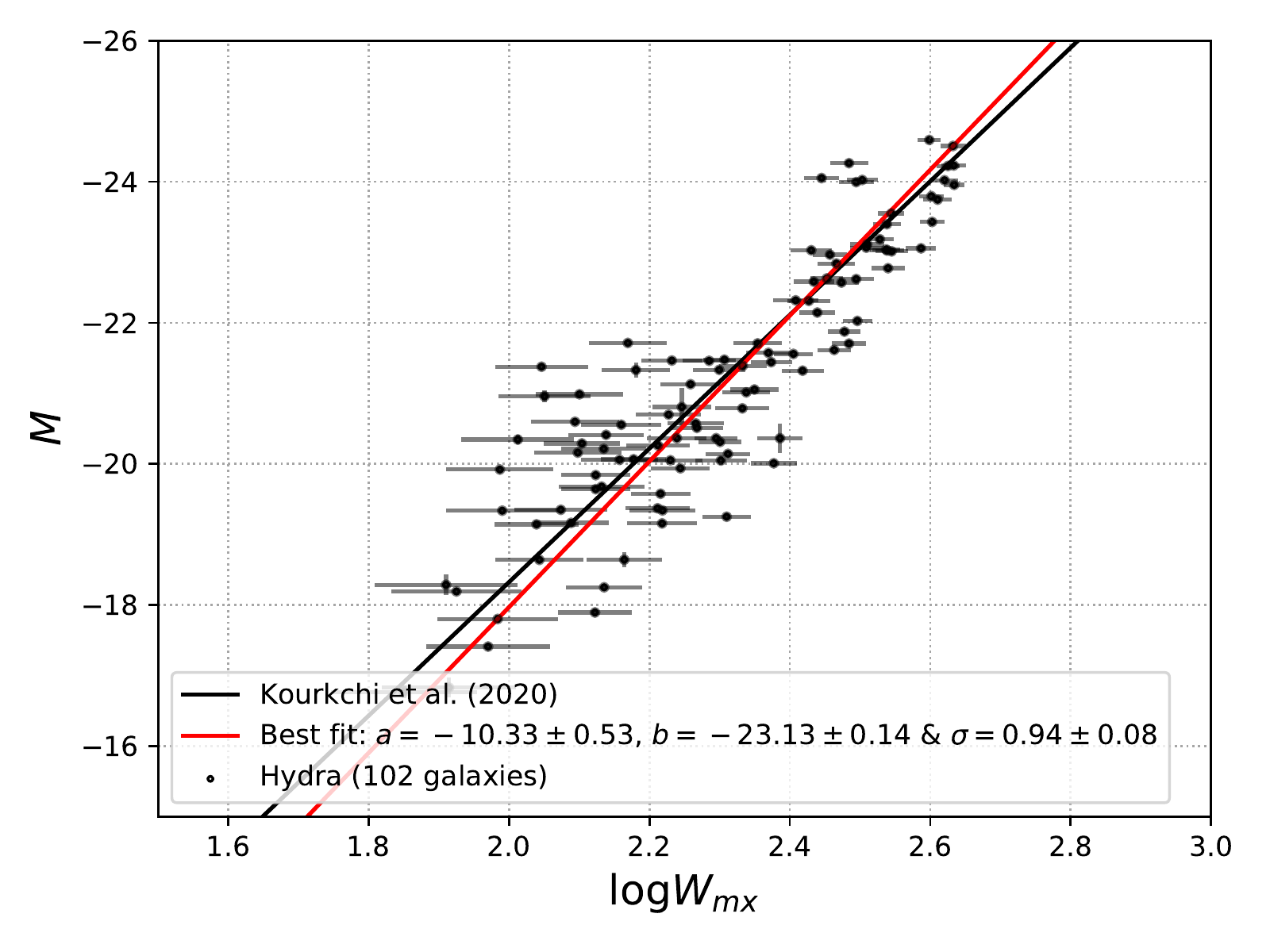}
\caption{Comparison of TFR for Hydra using the public WISE photometry (top-panel) and using our newly derived WISE photometry. The rms deviations from the best fit line (red-line) decreases from 3.6 to 0.9 when using the new photometry. Furthermore, the best fit parameters are very close to the universal calibrated relation. The best fit parameters (red lines) are from the hyperfit package \citep{2015PASA...32...33R} which uses errors in both axes.}
\label{Hydra_TF_old_vs_new}
%\end{center}
\end{figure}

We initially tried to use data from the WISE all sky catalog GATOR web interface. When plotting the TFR we found that this instrumental profile-fit magnitudes were not appropriate for photometry of extended sources. As can be seen in Fig.~\ref{Hydra_TF_old_vs_new}, the derived Tully-Fisher relation was totally unprobable.
%\textcolor{red}{(LSS comment: I suggest comparing the GATOR data with the new photometry, rather than plotting a meaningless TF relation. The approach of the previous section was to internally and externally validate velocity width data, before actually using it. The same should apply to the WISE photometry.)}

Thus for this analysis, the photometric parameters are measured from the WISE observations using the photometric pipeline from \cite{T_Jarrett2013}, where mosaics of each source are optimally constructed for resolved sources, carefully preserving the native angular resolution of WISE single frames \citep{TJarrett_2012}.  Source characterization includes size, shape, orientation, surface brightness and photometric measurements.  

These measurements serve as the basis for derived physical parameters, including size, luminosity, stellar mass and star formation rate. The aggregate stellar mass is estimated using the W1 in-band luminosity and the W1-W2 colour (to account for the population-dependent mass-to-light ratio), as described in \cite{MCluver_2014}.  The global star formation rate is estimated from the W3 and W4 spectral luminosities, after subtracting the stellar continuum, as detailed in \cite{MCluver_2017, TJarrett_2019}.
%he results includes measured parameters (e.g. apparent magnitude, radius, and axis ratio), derived photometric parameters (e.g. SED template, Mass-to-light ratio, and stellar mass).}

The pipeline returns several types of apparent magnitude including, 1-sigma isophotal magnitudes, asymptotic magnitudes, and total radial-profile-modeled magnitudes. The Tully-Fisher relation is an empirical relation between maximum rotation velocity and total luminosity, i.e total stellar mass) of the galaxy. We therefore used the total extrapolated magnitude which encompasses the isophotal measured magnitude plus extrapolation of the radial SB profile to 3 disk scale lengths for our analysis.

The second measured parameter that is needed for the TFR is the inclination which is extracted from the WISE observations . We used the measured axis ratio based on the W1 3-sigma isophote to derive the inclination.  We caution that the axis ratio and position angle are sensitive to the angular size of the galaxy due to the relatively large beam, FWHM = 6\arcsec, of the W1 imaging. Moreover, the near-infrared is sensitive to older stellar populations, and notably the bulge population, which tends to circularize the axis ratio and hence appear more face-on with inclination. 
%\textcolor{red}{(LSS comment: as in CF, is it possible to use data that's more sensitive to the disk population - e.g. SkyMapper/PanSTARRS?)}

% The photometric data are obtained from the Wide-field Infrared Survey Explorer \citep[WISE,][]{Wright:2010} All-Sky source catalog.  WISE 
% mapped the whole sky in four bands W1, W2, W3 and W4 centered at 3.4, 4.6, 12 and 22 $\mu$m, and provides more than 563 million measurements of
% infrared sources. The photometric quantities adopted in this paper are the instrumental profile-fit magnitudes and their uncertainties in the W1 
% band. The WALLABY spectrum sample was cross-matched with the WISE All-Sky source catalog using a search radius of 30 arcsec, which corresponds with the spatial resolution of WALLABY pilot survey.

% As for the disk inclination correction needed to de-project the HI linewidth, we have computed inclinations for all the galaxies using data provided by
%  \href{https://skymapper.anu.edu.au/}{SkyMapper data server}.
%\href{https://leda.univ-lyon1.fr}{Hyperleda database}.

\section{Tully-Fisher Distances}

\begin{figure*}
\begin{center}
\includegraphics[width=\textwidth,angle=-0]{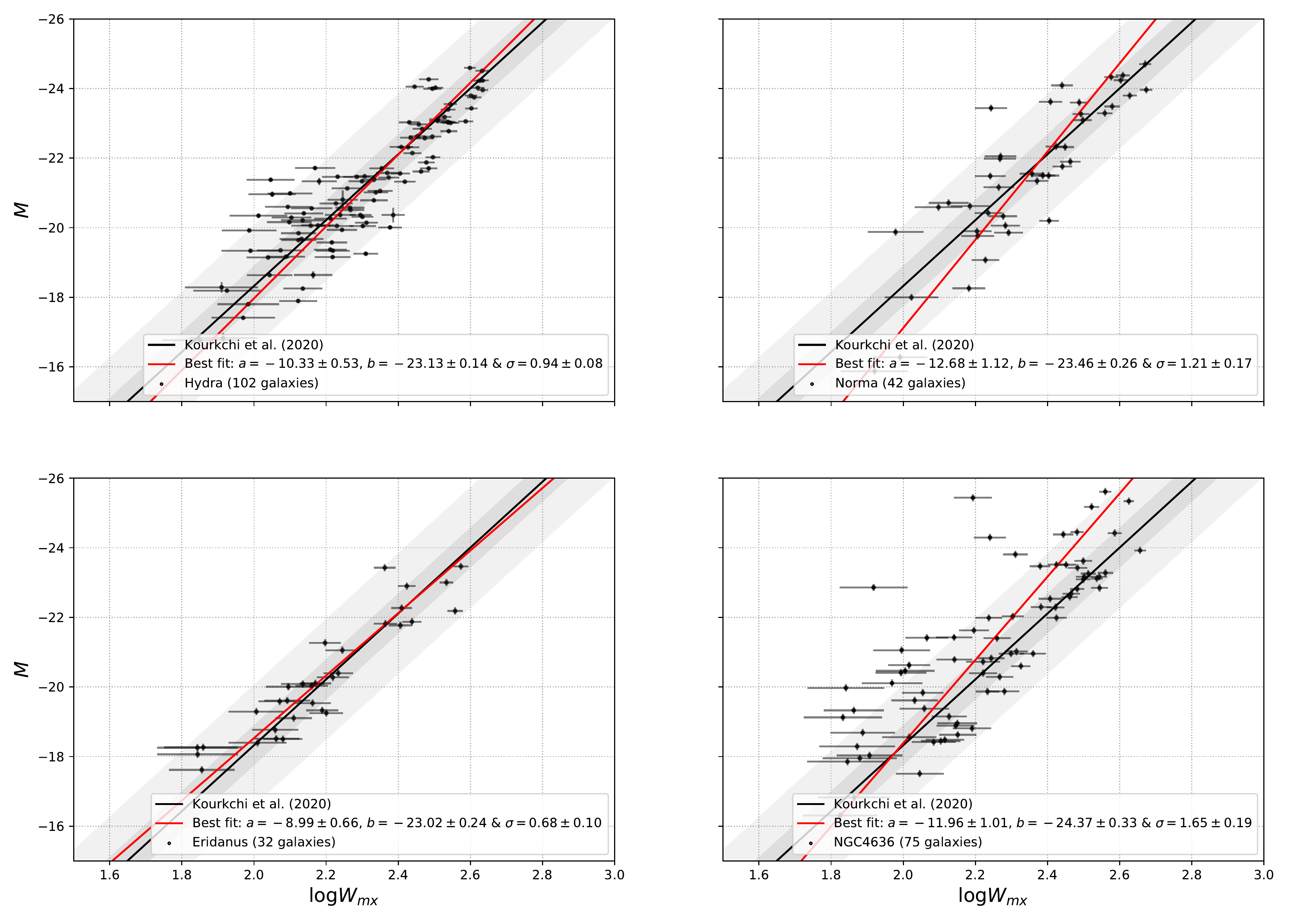}
\caption{TFR for the four fields in the WISE \textit{W1} band. The solid black line is the calibrated TFR from \protect\cite{Kourkchi:2020}; the dark and light shaded areas correspond to 1-$\sigma$ and 3-$\sigma$ scatter around the TFR, respectively; the solid red line is the best fit model. The quantities derived with the Pilot survey are in good agreement with the WISE \textit{W1} TFR calibration in K20 in the Hydra, Norma and Eridanus fields. The high scatter in the NGC4636 field is due to the noisy spectra caused by residual radio continuum.}
\label{fig:WALLABYtf}
\end{center}
\end{figure*}

Measuring distances requires a redshift independent distance indicator which can be either a primary indicator such as the Cepheid variable stars \citep{Fernie1969} or a secondary indicator such as, the Tully-Fisher relation \citep{Tully:1977}. In this paper, we will be adopting the Tully-Fisher relation as our workhorse to measure distances to spiral galaxies in the four WALLABY fields. As the name suggests, a secondary distance indicator such as the TFR should be calibrated first using a sample of spiral galaxies with known distances. 

\citet[K20 hereafter]{Kourkchi:2020} re-calibrated the TFR using optical \textit{u, g, r, i} and \textit{z} SDSS bands as well as infrared \textit{W1} and \textit{W2} WISE bands. In this article, we will use their WISE \textit{W1} calibrated relation as our standard tool in deriving the distances of galaxies in the WALLABY fields with, as plotted in Fig.~\ref{Hydra_TF_old_vs_new} and Fig.~\ref{fig:WALLABYtf}, a slope of $−9.47 \pm 0.14$, a zero point of $−20.36 \pm 0.07$ and rms of $0.58$. Although the sample used to re-calibrated the TF is different from our WALLABY sample, parameters, methods and corrections should be as consistent as possible. In this section we will go through all parameters that we used to derive the distances for all four WALLABY fields and their associated corrections.

With the calibrated TFR in WISE \textit{W1} (K20) in hand, two sets of observations are required to derive TF distances: (A) spectroscopic observations (in our case  HI 21-cm) from which we extract the galaxy rotational velocity; (B) photometric observations (in our case infrared imaging) from which we extract the apparent magnitude for each galaxy in our sample.

The first set of data is the WALLABY 21-cm spectra. The raw spectra were fitted using three different algorithms. The full presentation of this data set is in Section 2. To be consistent with the calibrated TFR from K20, we used the same method and corrections as  CosmicFlows to derive the $W_\mathrm{mx}$ line width (i.e., see section 2.2). The only difference is that, while CosmicFlows visually measured inclinations, here we are using WISE $W1$ band photometric parameters (Section 2.4) to derive the inclination as:
\begin{equation}
\text{cos}^2i = \frac{(b/a)^2-q_0^2}{1-q_0^2}
\label{inclination}
\end{equation}
where $q_0$ is the intrinsic axial ratio of the galaxy, $q_0$ = 0.13 for Sbc and Sc, and $q_0$ = 0.2 for other types \citep{Giovanelli1997}. We didn't make a search for the morphological type, so we used $q_0$ =0.2 if b/a $>$ 0.2 and $q_0$=0.13 if b/a $\le$ 0.2. Galaxies with axial ratio (b/a) greater or equal to 0.7 are excluded to avoid large corrections in the line width due to the inclination correction.

The heliocentric redshift can be measured from the WALLABY 21-cm spectra. 
We use redshifts to visualize the data (e.g., Fig.~\ref{fig:WALLABYtf} and \ref{fig:zWALLABY}). For the purpose of this paper, we first convert the heliocentric redshift $z_{helio}$ to the CMB redshift $z_{CMB}$ using,

\begin{equation}
z_\mathrm{CMB} = \frac{1+z_\mathrm{helio}}{1+z_\mathrm{Sun}} - 1~, 
\end{equation}
where $z_\mathrm{Sun}$ is the velocity of the sun with respect to the CMB in the direction of the galaxy being measured. It is calculated using the Planck dipole \citep{Planck2018}. We derive the comoving distance $D(z_\mathrm{CMB})$ in Mpc at redshift $z_{\mathrm{CMB}}$ using a flat $\Lambda$CDM with $H_0=75$ km s$^{-1}$ Mpc$^{-1}$ and $\Omega_m=0.27$ and convert this to luminosity distances using:
\begin{equation}
    D_L = (1+z_\mathrm{helio}) D(z_\mathrm{CMB})~.
\end{equation}

The second set of data consists of photometric imaging from WISE \textit{W1}. The full description of the data reduction process, from building mosaics to deriving the photometric parameters of interest, is given in section 2.4. The correction to apparent magnitudes is applied in the same manner as for the calibrated relation (K20):

\begin{equation}
    m_\mathrm{cor} = m_\mathrm{obs} - A_{W1} - k_{W1} - I_{W1}~,
    \label{appmag}
\end{equation}
where $m_{cor}$ is the desired corrected magnitude, $m_{obs}$ is the observed total magnitude, $A_{W1}$ is the extinction correction due to dust from our own Galaxy \citep{1998ApJ...500..525S}, $k_{W1}$ is the \textit{k-}correction of the form $A_{W1}^{k} = -2.27z$ \citep{1968ApJ...154...21O,2007ApJ...664..840H}, and  $I_{W1}$ is the correction for internal extinction which were considered but neglected because they are small in W1 band.  \cite{Sakai:2000} parameterize them as $\gamma \log a/b$. If the effective wavelength of W1 is 3.4$\mu$m, and the absorption in
magnitudes is proportional to $\lambda^{-1}$ , the maximum value is less than 0.1 mag for $\log W$ = 2.5.

The absolute magnitude for each galaxy, for the purpose of visualizing the data only, was calculated  as:
\begin{equation}
    M = m_\mathrm{cor} - 5\log_{10}(D_L) - 25~.
    \label{absMag}
\end{equation}

Figure \ref{fig:WALLABYtf} visualizes the three ingredients needed for the TFR: (A) the solid line denotes the calibrated TFR in WISE \textit{W1} and the associated 1-$\sigma$ and 3-$\sigma$ regions (shaded areas); (B) the velocity width $\log W_\mathrm{mx}$, on the $x$-axis is the maximum rotational velocity derived from the WALLABY spectra (Section 2.2); (C) the absolute magnitude $M$, on the $y$-axis is derived from Equation \ref{absMag}. The TFR fit is performed using the hyperfit method \citep{2015PASA...32...33R}, allowing varying uncertainties in both axes.

We calculated the scatter about the TFR for the four WALLABY fields in Fig.~\ref{fig:WALLABYtf}. The rms for Hydra is $0.90\pm0.08$ mag, which is comparable to the value of $0.89$ found by K20 for the same cluster. The rms scatter in the Norma field is $1.18\pm0.17$ mag which is slightly higher than most of the individual clusters in K20. The higher rms scatter in the TFR in Norma field is due to the strong dust extinction and stellar confusion which affects the photometric data reduction. The scatter about the TFR in Eridanus field is $0.58\pm0.08$ mag which is the smallest of all the four fields and is identical to the rms scatter of the final calibrated TFR from K20. The rms scatter about the TFR in the NGC4636 region is the highest at $1.61\pm0.19$ mag. The source of this scatter may be the noisy WALLABY spectra with baseline ripples due to residuals caused by strong radio continuum sources in the NGC4636 field (see Fig.~\ref{fig:compwmx}). 

Furthermore, source finding in the N4636 group was done in a different way from the other fields  (Hydra, Norma and Eridanus). It was done by searching at the coordinates (RA, Dec, $z$) of previously catalogued galaxies, while the source finding for the other clusters was done blindly. The master catalogue providing input coordinates for NGC4636 was a combination of SDSS, ALFALFA, HIPASS and 6dFGS. 
%Because ALFALFA does not reach south, and HIPASS is much shallower than ALFALFA and WALLABY, the ASKAP HI detection in NGC4636 may be less complete than in other Pre-pilot and Pilot WALLABY fields.
%LSS: catalogues are mostly optical
Moreover the northern tile of the NGC 4636 field consists of only a single footprint, as the other footprint had to be discarded due to data quality issues. Hence, the noise level in the northern half of the field is significantly higher than in all other fields.

With all parameters in hand, the absolute magnitude of each galaxy is calculated in this article using the slope and zero point provided in K20 given its maximum rotational velocity :

\begin{equation}
    M(W_\mathrm{mx}) = -(20.36\pm0.07) - (9.47\pm0.14)(\log_{10}W_{mx} - 2.5) - 2.699~,
    \label{absMw}
\end{equation}
where  the TFR slope is ($-9.47\pm0.14$), the zero point is ($-20.36\pm0.07$), and the conversion of the AB to the Vega system is 2.699. 

Using the apparent and absolute magnitude derived from equations \ref{appmag} \& \ref{absMw}, we can calculate the distance modulus as:

\begin{equation}
    \mu = m_\mathrm{cor} - M(W_\mathrm{mx}) .
\end{equation}

\begin{figure}
%\begin{center}
\includegraphics[width=\columnwidth,angle=-0]{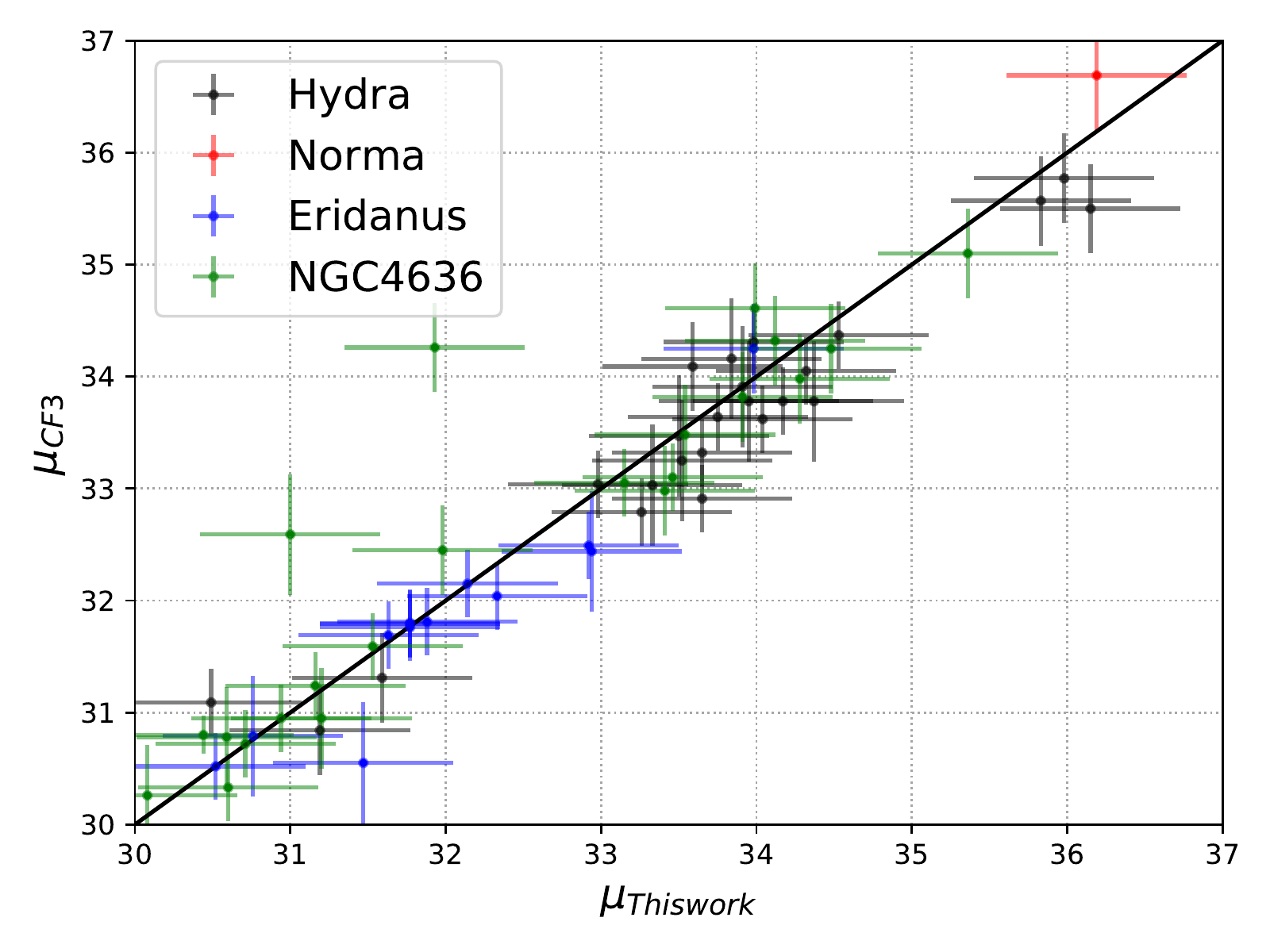}
\caption{Comparison of the distance modulus measurements from the WALLABY pre-pilot survey (this work) vs. CosmicFlows-3 \protect\citep{Tully2016}}
\label{fig:WALLABYvscf3}
\end{figure}

Cross-matching with the CosmicFlows-3 (CF3) catalog \citep{Tully2016} result in 24 source matches in Hydra, 1 in Norma, 13 in Eridanus, and 24 in the NGC4636 field. We compared our measurements for the distance modulus to the published ones from CF3. Figure \ref{fig:WALLABYvscf3} shows the comparison of the distance modulus from the four WALLABY fields with the ones from CF3. Figure \ref{fig:WALLABYvscf3} shows that our measurements appear to be systematically higher than the CF3 distance moduli  measurements. This can be due to the value of the zero point calibration which was adjusted for the CF3 measurements specifically, but not for our WALLABY measurements. The zero point calibration will not be an issue for WALLABY measurements in the future, given the large sky-area which will allow it to easily be adjusted using more accurate measurements such as SN Ia.

\begin{figure}
%\begin{center}
\includegraphics[width=0.5\textwidth,angle=-0]{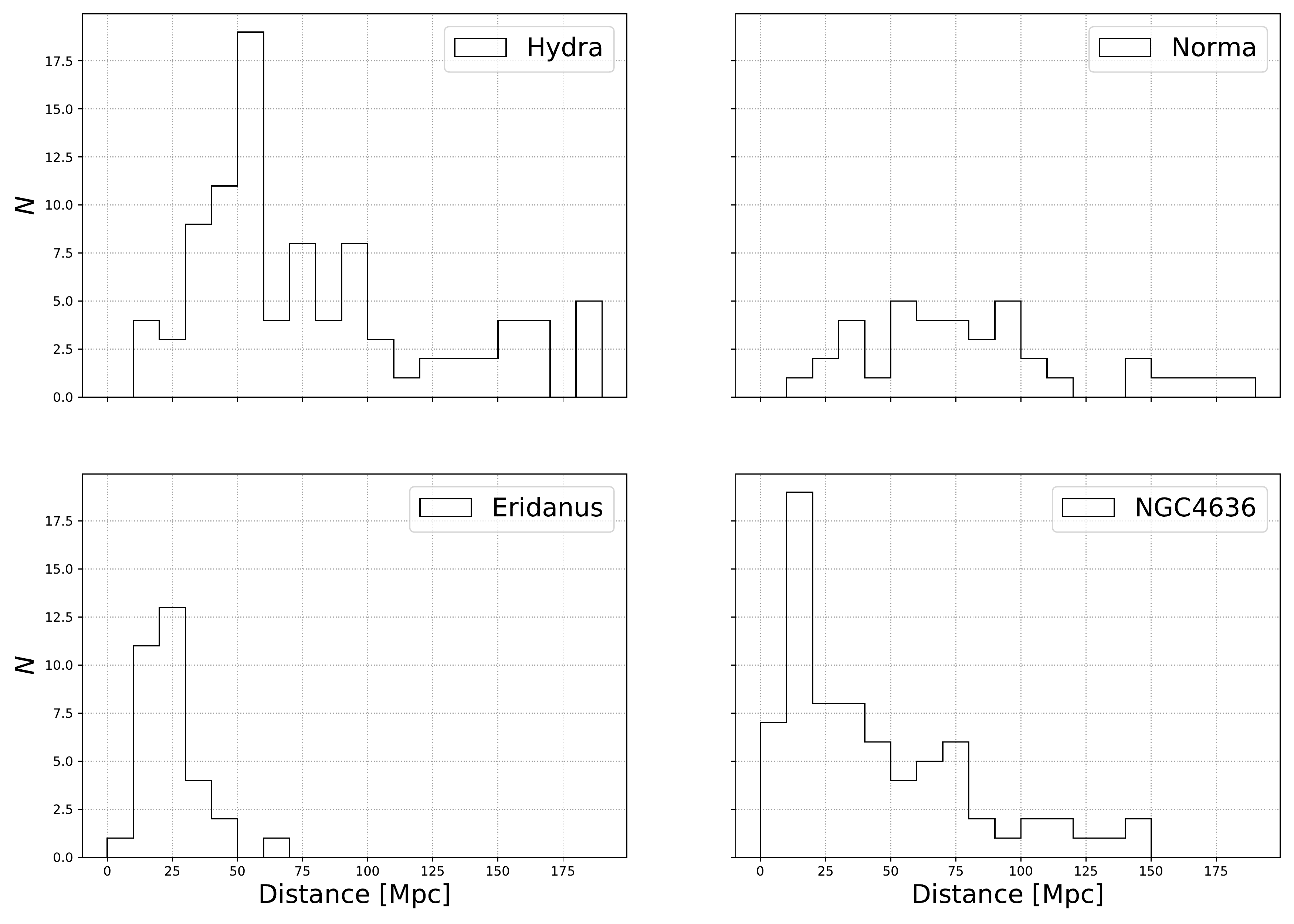}
\caption{TFR distance distribution of galaxies in the four fields after extinction and k-corrections and the b/a cut $<$ 0.7 using K20 calibration.}
\label{fig:dWALLABY}
\end{figure}

The distribution of the measured TF distances in the four WALLABY fields are shown in Fig.~\ref{fig:dWALLABY}. The WALLABY Hydra field includes background galaxies that do not belong to the  Hydra cluster. The same applies to the Norma and Eridanus fields. Measuring distances to the Hydra, Norma, NGC 4636 and Eridanus clusters will therefore require restrictions on redshift. 

% \begin{figure}
% \begin{center}
% \includegraphics[width=\columnwidth]{LEGACY/TF_Hydra_old_wise.pdf}
% \includegraphics[width=\columnwidth]{LEGACY/TF_Hydra_new_wise.pdf}
% \caption{Comparison of TFR for Hydra using the public WISE Photometry (top-panel) and using our newly derived WISE Photometry. The rms deviations from the best fit line (red-line) decreases from 3.6 to 0.9 when using the new photometry. Furthermore, the best fit parameters are very close to the universal calibrated relation. The best fit parameters (red-lines) uses the Hyperfit package (Howlett et al. in prep) which adopt errors in both axes.}
% \label{Hydra_TF_old_vs_new}
% \end{center}
% \end{figure}

%\subsection{The Hydra and Norma clusters}
Fortunately, there is a clear distinction in the redshift distribution between the galaxies in the cluster and the background galaxies in the Hydra, Norma, and Eridanus fields. Figure \ref{fig:zWALLABY} shows the redshift distribution in the four WALLABY fields. The dashed-red lines represent the distinction line between the cluster galaxies and the background ones. Using only galaxies with redshifts in the low-redshift side of the dashed-red lines, we measured the weighted average distance to Hydra, Norma, NGC 4636 and Eridanus to be $53.5\pm2.5$, $69.4\pm5.5$, $23.0 \pm 2.1$  and $21.5\pm1.4$ Mpc, respectively. In this study, the Norma distance refers to the eastern overdensity (which is part of the Norma supercluster). This distance may be slightly different from the cluster distance due to filament projection.
In the next section a comparison with previous determinations of these distances will be made.

Peculiar velocities are the non-Hubble velocities induced gravitationally by large scale
structure. For a Hubble Constant between 68 and 73 km/s/Mpc the peculiar velocities with respect to the CMB of the
Hydra, Norma, NGC 4636 and Eridanus clusters lie in the range (--320, --40), (--730, --350),
(300, 430) and (0, 110) km s$^{-1}$ respectively.

\section{Spatial distribution of galaxies in the Phase 1 Pilot Survey}

\begin{figure}
\begin{center}
\includegraphics[width=0.5\textwidth,angle=-0]{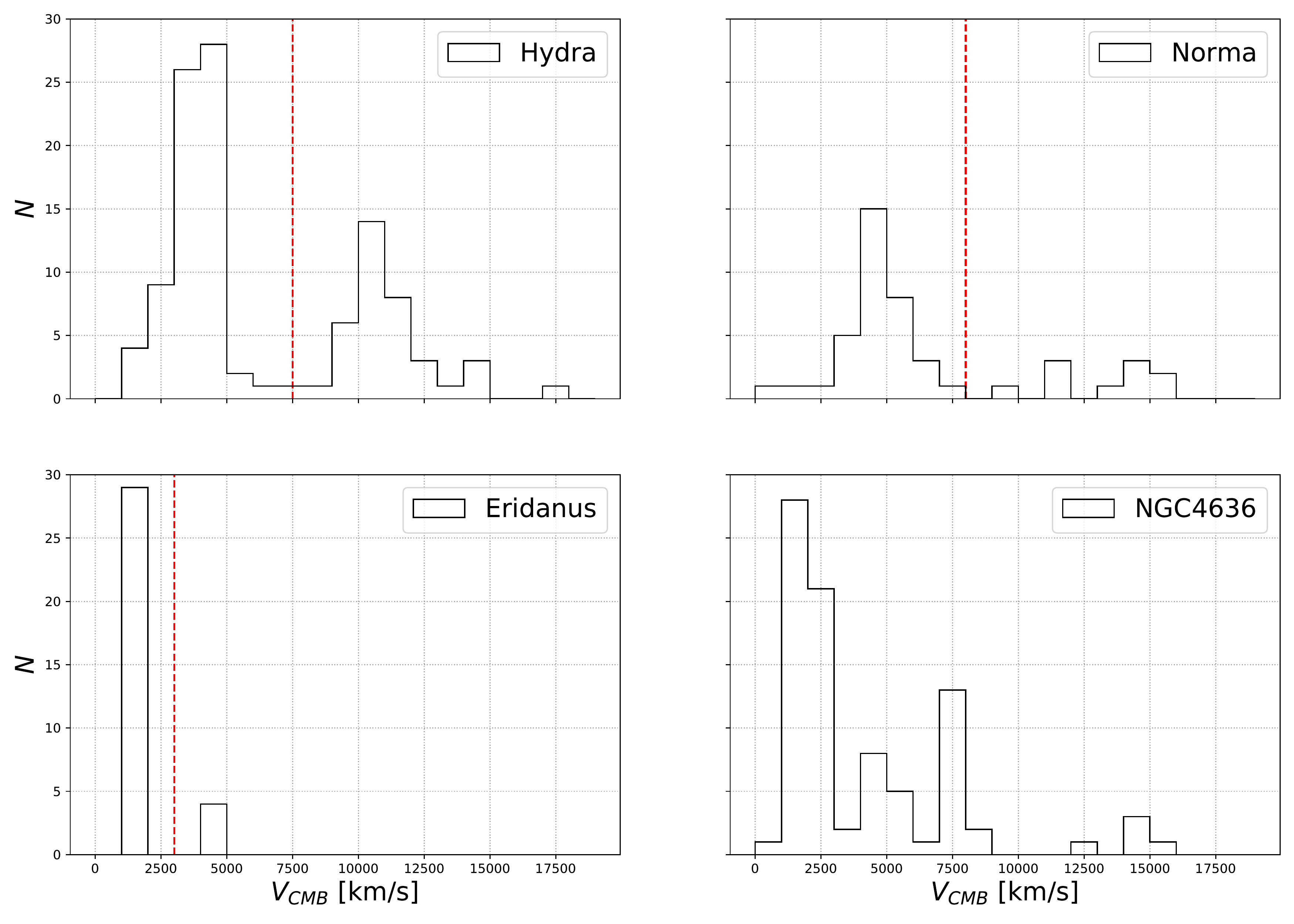}
\caption{The redshift distribution (CMB frame) of all 614 galaxies in the Hydra, Norma, Eridanus and NGC4636 fields. The dashed-red lines in the Hydra, Norma and Eridanus plots represent the estimated division between the cluster  and background galaxies. The WALLABY pilot integration time allows the detection of galaxies located at 15,000 km s$^{-1}$, around twice the distance limit of previous single dish surveys such as HIPASS.} 
\label{fig:zWALLABY}
\end{center}
\end{figure}

Figure \ref{fig:zWALLABY} illustrates the variance in the redshift distribution between the WALLABY pilot survey fields and the presence of cluster foreground and background galaxies. It also shows that the pilot integration times allow us to reach a recessional velocity of 15,000 km~s$^{-1}$, doubling what has been achieved by predecessor all-sky surveys reliant on single dish observations, such as 2MTF. These higher redshifts have so far only been accessible using fundamental plane techniques -- e.g.\  6dFGS, which provides about 6,000 useful peculiar velocities. WALLABY is expected to deliver about 30,000-50,000 peculiar velocities.

\subsection{Hydra and Norma cluster fields}

\begin{figure*}
\begin{center}
\includegraphics[width=\textwidth,angle=-0]{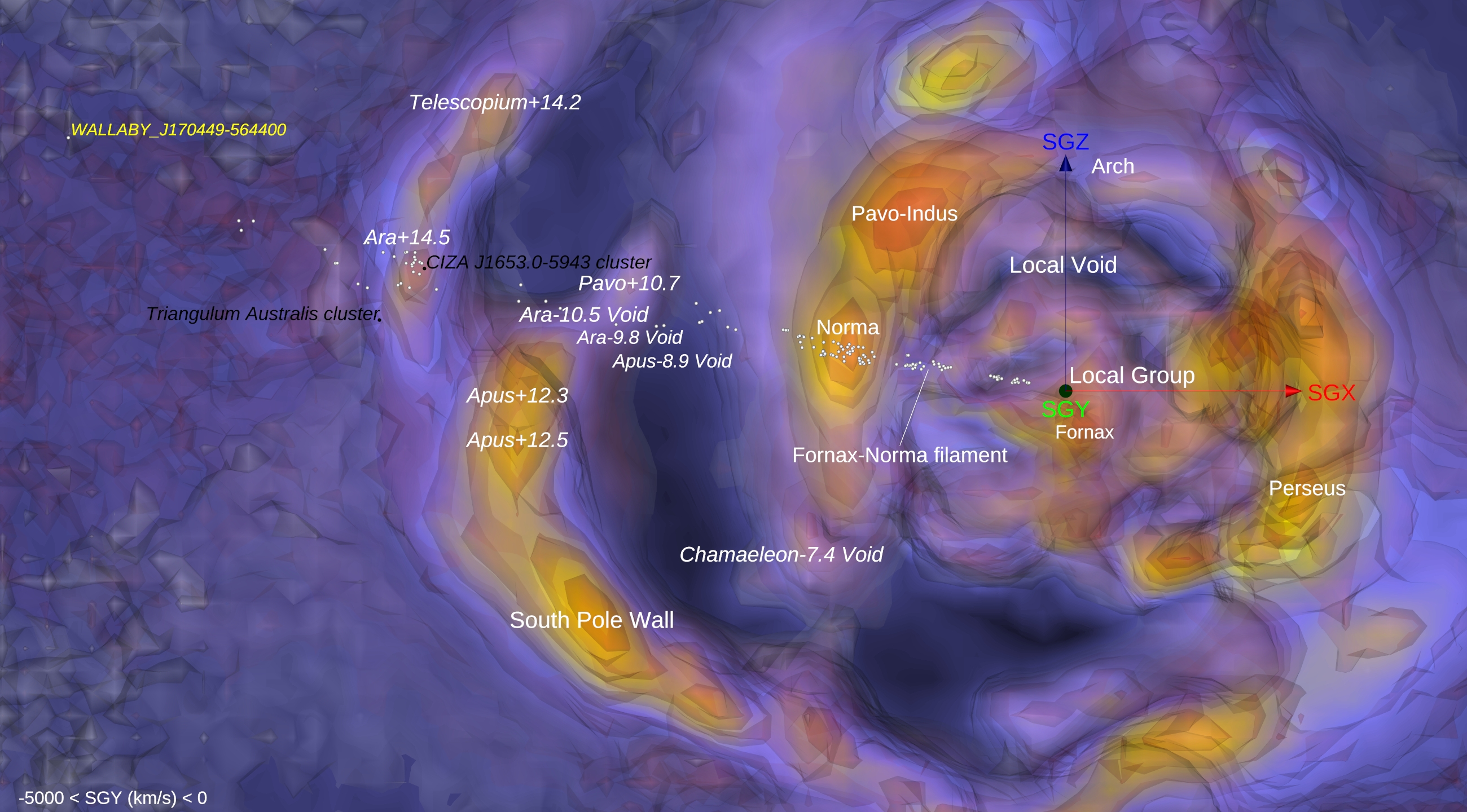}
\caption{Map of the redshift positions of galaxies in the Norma field (white points) overlaid on the (coloured) density contrast field ($\delta$) as reconstructed from the CosmicFlows-3 Catalog of peculiar velocities.
Scale and orientation are provided by the red and blue 5000 km s$^{-1}$-long arrows emanating from our position and associated with the supergalactic coordinates SGX and SGY, respectively. A slice defined by $0>$SGY$>-5000$ km~s$^{-1}$ is here seen from the negative SGY direction. This static view is complemented by an online \href{https://sketchfab.com/3d-models/WALLABY-norma-vs-cf3-density-e41a006e46cc44b69c6e656c000bbec2}{interactive 3D visualization of the Norma survey field} showing three levels of the overdensity $\delta$, field. This tool can be used to better grasp how the field includes a series of voids and overdensities.}
\label{fig:map_norma} 
\end{center}
\end{figure*}

The Hydra cluster has a mean redshift of 0.012 \citep{BabykVavilova:2013}
and a velocity dispersion of 690 km s$^{-1}$ \citep{Lima-Dias:2021}. Figure \ref{fig:zWALLABY}  shows
the redshift distribution 
for Hydra cluster galaxies in our sample 
%with velocities in column (2) of Table 1 between 1700 and 6000 km s$^{-1}$ec. 
whereas Fig.~\ref{fig:WALLABYtf} shows the TFR for these galaxies.

\begin{figure*}
\begin{center}
\includegraphics[width=\textwidth,angle=-0]{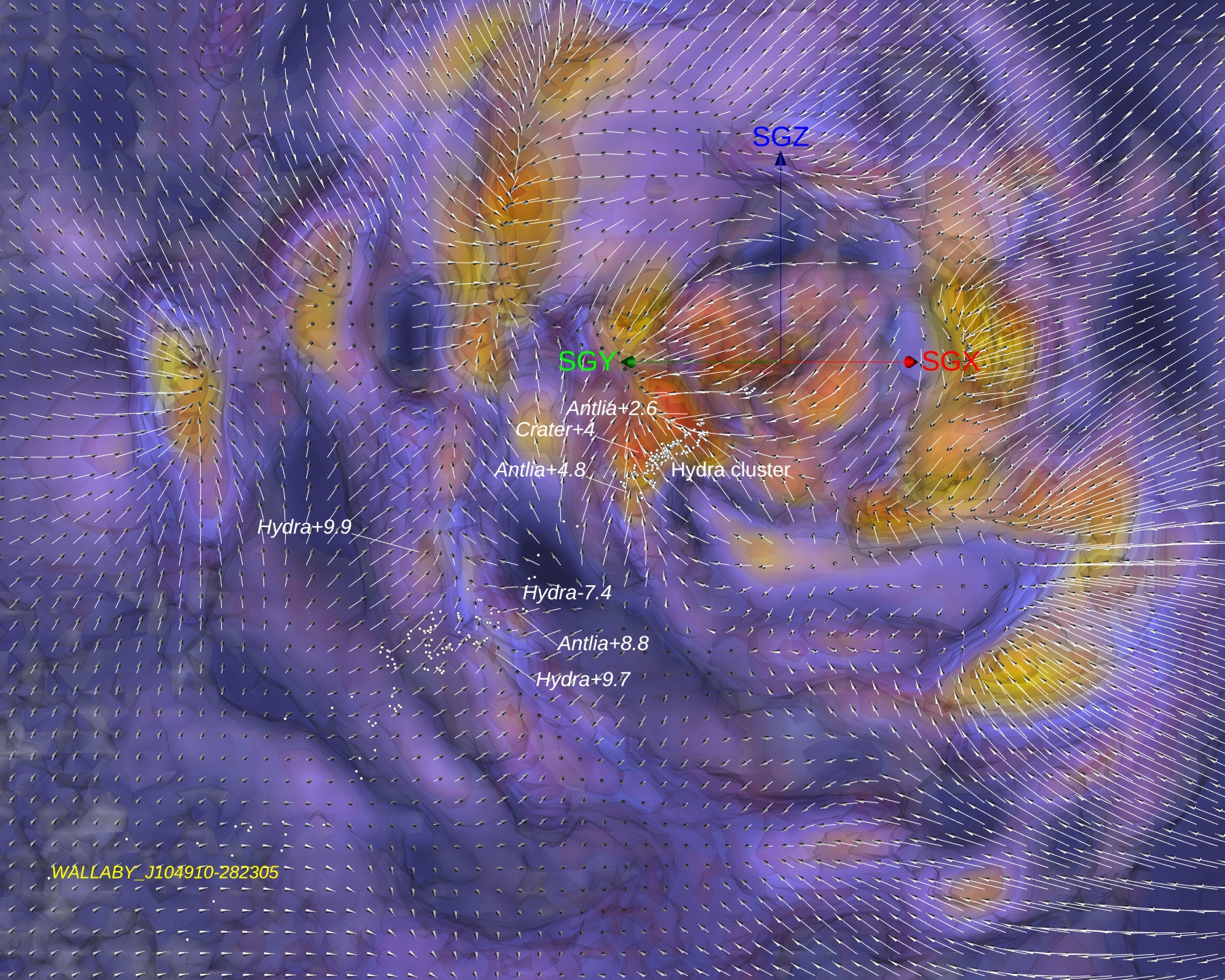}
\caption{Map of the redshift positions of galaxies in the Hydra field (white points) overlaid on the (coloured) full matter (dark + luminous) density contrast $\delta$, field and the peculiar (gravitational) velocity field (white arrows) as reconstructed from the CosmicFlows-3 Catalog.
Scale and orientation are provided by the red, green and blue 5000 km s$^{-1}$-long arrows emanating from our position and associated with the three cardinal axes of the supergalactic coordinate system SGX, SGY, and SGZ, respectively. A slice of 5000 kms$^{-1}$-thickness centered on the plane defined by the supergalactic longitude of the Hydra cluster (SGL=$139.36^\circ$) is here seen face-on. This static view is complemented by an online \href{https://sketchfab.com/3d-models/WALLABY-hydra-vs-cf3-density-9607eeb2df0c42859c3b1499b03c9ae9}{interactive 3D visualization of the Hydra survey beam} showing three levels of the overdensity $\delta$, field. This tool can be used to better grasp how the field includes a series of voids and overdensities.}
\label{fig:map_hydra} 
\end{center}
\end{figure*}

\cite{Woudt:2008} find a mean velocity of 4871 $\pm$ 54 km s$^{-1}$ for the Norma
cluster and locate its centre at ($l,b$) = (325.3$^\circ$, --7.2$^\circ$). They
identify it as the richest cluster in the region of the Great Attractor.
\cite{Mutabazi:2021}, in a study of its fundamental plane, finds a small peculiar velocity. The WALLABY pilot survey field is centred on ($l,b$) = (329.3$^\circ$, --10.4$^\circ$), i.e. 5.1$^\circ$, respectively 6 Mpc from the cluster center. Nevertheless, there is
clear overdensity of galaxies visible at the cluster redshift (Fig.~\ref{fig:zWALLABY}).\\

A full 3D view of the Great Attractor region  with the first CosmicFlows reconstruction of the local velocity field near Norma and Hydra clusters was presented by \cite{Courtois:2012}. Infall to the Norma cluster takes the form of a sharper distribution in redshift space than in real space (see Fig.~\ref{fig:map_norma}).

The positions of WALLABY galaxies within the Cosmic Web can be studied using the full matter (dark and luminous) density contrast field (usually noted $\delta$) reconstructed from the CosmicFlows-3 Catalog of peculiar velocities \citep{Graziani:2019,2019ApJ...880...24T,2019MNRAS.490L..57C,2020ApJ...897..133P}.
Figure~\ref{fig:map_norma} shows the redshift locations of galaxies of the Norma survey versus a real-space three-dimensional visualization of the overdensity field, restricted to a slice. The visualization is obtained by a combination of a ray-casting algorithm \citep{2017PASP..129e8002P} resulting in a smooth rendering of the $\delta$ overdensity field ranging from most underdense (deep blue colour) to most overdense (yellow), and a series of semi-transparent isosurface polygons resulting in a sharp materialization of the surfaces ranging from $\delta=0$ (grey surface) to highly overdense ($\delta=2.8$ in red). The grey surface hence marks the frontier between the underdense and the overdense Universe. The most noticeable feature in the distribution of WALLABY galaxies is their aggregation near the Norma cluster at SGX $\simeq$ $-5000$ km s$^{-1}$, seen as a pronounced local maximum in the density contrast field ($\delta$ field). Between the Local Group and the Norma cluster, the field includes a section of the Local Void \citep{2019ApJ...880...24T}  and the termination of a filament connecting the Fornax cluster to the Norma cluster, as can better be seen from the interactive 3D visualization supplementing Fig.~\ref{fig:map_norma} (see annotations 2 and 3). Beyond the Norma-Pavo-Indus filament \citep{Fairall,2013AJ....146...69C}, the field  includes the void separating this filament and the South Pole Wall \citep{2020ApJ...897..133P}. The map of Fig.~\ref{fig:map_norma} indicates the location of local extrema of the over and under-density field $\delta$, such as the Ara-10.5 Void and the Ara+14.5 overdensity, whose names include the name of the constellation in which they sit, followed by a `+' for an overdensity or a `$-$' for an underdensity, followed by their redshift in units of 1000 km~s$^{-1}$, a convention established in \cite{2019ApJ...880...24T}. An aggregation of WALLABY galaxies is found at the location of the Ara+14.5 knot of the Cosmic Web, that is nearly coincident with the CIZA J1653.0-5943 cluster \citep{2002ApJ...580..774E}. Beyond this most distant section of the South Pole Wall, the field presumably enters a new void, in regions where the CosmicFlows-3 reconstruction tends to a null field due lack of sufficient data. The most distant galaxy in the Norma field, located at redshift 0.073, is indicated on the map.

In Fig.~\ref{fig:map_hydra} the distribution of the WALLABY galaxies in the Hydra field is plotted in a similar way on top of the CosmicFlows-3 cosmography. The density of WALLABY galaxies as a function of distance displays contrasting features: after displaying a handful of galaxies located in our vicinity at $\simeq$ 900 km s$^{-1}$, no galaxies are detected in the void located in the foreground of the Hydra cluster. A major aggregation of objects is then seen at the crossing of the Hydra cluster, which is part of a high-density node in the Cosmic Web, as seen in the map of the CosmicFlows-3 density contrast. In the background of the Hydra cluster, the surveyed region includes a deep void (Hydra-7.4) where very few galaxies reside. In the background of this void, there is an extended cloud of galaxies scattered throughout a wall located in Hydra at about 10,000 km s$^{-1}$, where local maxima in the density are found: Hydra+9.7, Hydra+9.9 and Antlia+8.8. After dropping as a function of distance, the number density of galaxy rises again at the most extreme distances, with WALLABY\, J104910-282305 being the most remote one at $\sim$ 23,000 km s$^{-1}$. This infall pattern
of peculiar velocities leads to a higher overdensity in redshift space (Fig.~\ref{fig:zWALLABY})
than in real space (Fig.~\ref{fig:dWALLABY}).

\subsection{The NGC 4636 cluster and Eridanus fields}
\begin{figure*}
\begin{center}
\includegraphics[width=\textwidth,angle=-0]{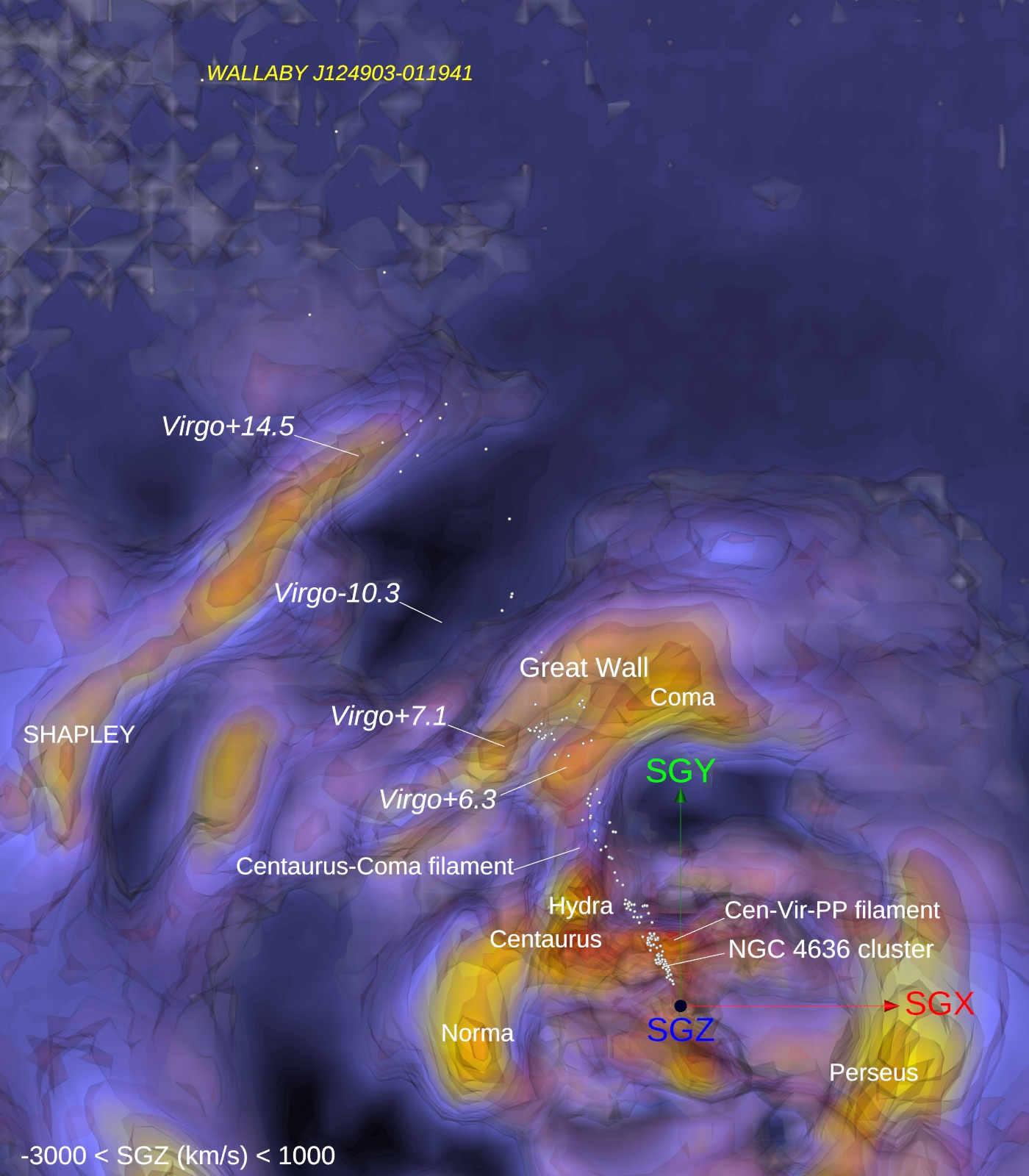}
\caption{Map of the redshift positions of galaxies in the NGC 4636 field (white points) overlaid on the (coloured) density contrast $\delta$, as reconstructed from the CosmicFlows-3 Catalog of peculiar velocities.
Scale and orientation are provided by the red and green 5000 km~s$^{-1}$-long arrows emanating from our position and associated with supergalactic coordinates SGX and SGY, respectively. A slice contained defined by $1000>$SGZ$>-3000$ km~s$^{-1}$ is here seen from the positive SGZ direction. This static view is complemented by an online \href{https://sketchfab.com/3d-models/cf3-vs-WALLABY-ps-ngc4636-dr1-catalog-69f02558e3a64e98a14a03d3e845f75b}{interactive 3D visualization of the NGC 4636 survey field}.}
\label{fig:map_ngc4636} 
\end{center}
\end{figure*}

The spatial distribution of the galaxies of the NGC 4636 survey is shown in Fig.~\ref{fig:map_ngc4636}. The  highest density in the field is at the crossing of the Virgo Strand \citep{2013AJ....146...69C}, a section of the Cen-Vir-PP filament that connects the Centaurus and Perseus nodes of the Cosmic Web through the Virgo cluster \citep{2017ApJ...845...55P}, in which the NGC 4636 cluster is nested. Beyond the cluster, the field includes the outskirts of the void located between Virgo and the Great Wall \citep{1978ApJ...222..784G}, and also includes part of the Centaurus-Coma filament \citep{2017ApJ...845...55P}. The number of galaxies then rises at the intersection of the Great Wall \citep{1986ApJ...302L...1D} near the Virgo+6.3 and Virgo+7.1 local maxima of the density contrast. In the background of the Great Wall, there is the trough and peak of Virgo-10.3 and Virgo+14.5.

\begin{figure*}
\begin{center}
\includegraphics[width=\textwidth,angle=-0]{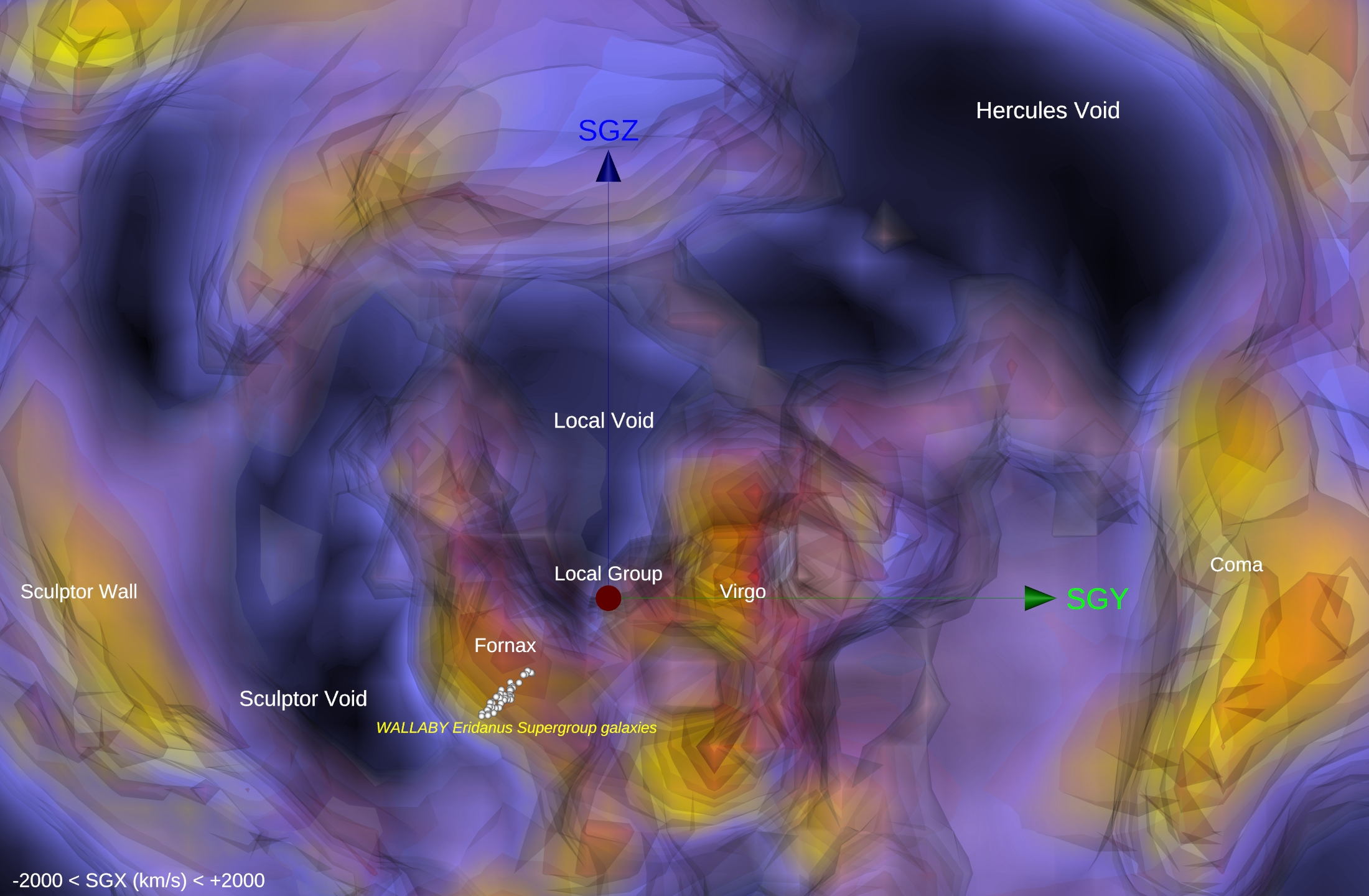}
\caption{Map of the redshift positions of galaxies in the Eridanus super group (white points) overlaid on the (coloured) density contrast $\delta$, as reconstructed from the CosmicFlows-3 Catalog of peculiar velocities.
Scale and orientation are provided by green and blue 5000 km s$^{-1}$-long arrows emanating from our position and associated with supergalactic coordinates SGY and SGZ, respectively. A slice defined by $2000>$SGX$>-2000$ km~s$^{-1}$ is here seen from the positive SGX direction. This static view is complemented by an online \href{https://sketchfab.com/3d-models/WALLABY-eridanus-supergroup-vs-cf3-density-4e4c7301669a4c55b295e2d781f4b308}{interactive 3D visualization of the Eridanus super group survey}.}
\label{fig:map_eridanus} 
\end{center}
\end{figure*}

Finally, the location of the galaxies of the Eridanus super group survey is presented in Fig.~\ref{fig:map_eridanus}. It shows that the galaxies are found near the Fornax node in the Cosmic Web, and distributed along the density gradient seen in the transition toward the Sculptor Void located in the background \citep{1990AJ.....99..751P,Fairall}.

\subsection{Discussion}

The cosmographic maps presented in the previous section demonstrate that the WALLABY pilot survey fields probe a vast diversity of elements of the cosmic web: intersecting numerous features such as voids, filaments, walls, and the nodes which host the targeted clusters. The future inclusion of WALLABY-detected galaxies in the CosmicFlows collection will add useful data for cosmographical purposes. Of particular interest are the regions around the Hydra and Norma clusters and their roles in the dynamical structure and influence of the Great Attractor \citep{1987ApJ...313L..37D,Woudt:2008,2014Natur.513...71T,2017NatAs...1E..36H}. WALLABY has the advantage of being able to peer through the Galactic Plane, permitting the study of the so-called Zone of Avoidance which  limits optical/IR surveys. Also of great interest are galaxies that reside in voids, which will provide improved constraints on the flows evacuating these underdense patches \citep{2008ApJ...676..184T,2017ApJ...835...78R,2017ApJ...850..207S}.
The most distant sections of the pilot fields, with velocities out to 22,000 km s$^{-1}$, will bring in new useful information on the architecture of the Cosmic Web in remote locations, thus providing better measurements of the large-scale gravitational field than currently available and potentially providing a definitive explanation for the amplitude of the CMB dipole.

In the vicinity of the Great Attractor, peculiar velocities for the Hydra and Norma fields show an infall pattern towards the respective clusters, i.e. positive peculiar velocities on the nearside of the clusters and negative on the far side. Beyond Hydra, positive
velocities return as the galaxies come under the influence of the wall located at about 10,000 km s$^{-1}$ $\simeq$ 135 Mpc, seen in Fig.~\ref{fig:map_hydra} at the locations of the local maxima Hydra+9.7 and Hydra+9.9. The pattern observed in the peculiar velocities at $\sim$ 100 Mpc is typical of an evacuation by the void located at 7400 km s$^{-1}$, labelled Hydra-7.4 in Fig.~\ref{fig:map_hydra}, where galaxies located in the foreground of this void have negative velocities, while galaxies located in the background have positive velocities. The wealth of data represented by the WALLABY galaxies located in this direction at large distance will play an important role in mapping the structure currently hinted at by the CosmicFlows-3 reconstruction, in particular of the wall seen in Hydra at 10,000 km s$^{-1}$. 

\section{Conclusions}
The Tully-Fisher distance of the Eridanus group is 21.5 $\pm$ 1.4 Mpc, and its peculiar velocity for a Hubble constant of 68--73 km s$^{-1}$ Mpc$^{-1}$ is small, between 0 and 110 km s$^{-1}$.
The NGC 4636 group is a complex structure within the Virgo cluster zero velocity surface,
and its Tully-Fisher distance is 23.0 $\pm$ 2.1 Mpc. Under the same assumptions its
peculiar velocity is 300--430 km s$^{-1}$. The distance of the Hydra cluster is 53,5 $\pm$
2.5 Mpc. There is a substantial background group associated with the Shapley Supercluster.
We measure a peculiar velocity of --320 to --40 km s$^{-1}$ for the Hydra cluster,
assuming 68 $<$ H$_0$ $<$ 73 km s$^{-1}$ Mpc$^{-1}$. The Norma cluster is the most distant in the
present sample at 69.4 $\pm$ 5,5 Mpc. Its peculiar velocity is --350 to --730 km s$^{-1}$
under the same assumptions.

This study also aims at testing the reliability, accuracy and dynamic range in distance able to be provided by  Tully-Fisher distances for galaxies detected and parameterised in the pre-pilot and Phase 1 surveys of WALLABY. It is also a prediction for what the full survey will deliver. We find that :
\begin{itemize}
\item Measurements of galaxy velocity widths are in excellent agreement with those measured using
previous single-dish observations with typical discrepancies around 7 km s$^{-1}$, validating both the data quality and observational strategy of WALLABY. The WALLABY Tully-Fisher distances have uncertainties at the level of 5 to 10\% which are similar or better than the accuracy obtained with single dish observations.
\item Even as close to the Zone of Avoidance as the Norma cluster, HI spectra provided by the SoFiA pipeline are excellent and extinction in the WISE W1 band is less than 0.1 mag. The combination of WALLABY and WISE W1 band observations will enable investigation of large number of accurate peculiar velocities of galaxies
in the upcoming full WALLABY survey. The Tully-Fisher dispersion obtained is at the level of 0.9 unit which is excellent for these cluster fields.
\item The distribution of galaxies and kinematic pattern in the Hydra and Norma fields is one of infall to the cluster mass concentrations.
\item The WALLABY pilot survey sensitivity and widefield capability will extend the investigation
of peculiar velocities beyond what was previously achieved with the single dish TFR method. WALLABY will detect about 500k galaxies at a mean redshift of $z = 0.05$. If 251 out of 614 of the pilot phase 1 galaxies pass the S/N and inclination tests, this might suggest that 200k Tully-Fisher distances would result from WALLABY – much higher than was expected in initial proposals and increasing by a factor of twenty the current number of Tully-Fisher distances in Cosmic-Flows-4. The number may be slightly optimistic given that we only observed nearby structures on Phase 1. However, most fields should have their peak redshift at 0.04, well beyond these local structures.

\end{itemize}

%%%%%%%%%%%%%%%%%%%%%%%%%%%%%%%%%%%%%%%%%%%%%%%%%%
\section*{Acknowledgements}
The Australian SKA Pathfinder is part of the Australia Telescope National Facility which is managed by CSIRO. Operation of ASKAP is funded by the Australian Government with support from the National Collaborative Research Infrastructure Strategy. 
ASKAP uses the resources of the Pawsey Supercomputing Centre. Establishment of ASKAP, the Murchison Radio-astronomy Observatory and the Pawsey Supercomputing Centre are initiatives of the Australian Government, with support from the Government of Western Australia and the Science and Industry Endowment Fund. 
We acknowledge the Wajarri Yamatji people as the traditional owners of the Observatory site. Parts of this research were supported by the Australian Research Council Centre of Excellence for All Sky Astrophysics in 3 Dimensions (ASTRO 3D), through project number CE170100013. We would like to thank all our colleagues on the WALLABY team for helpful discussions.  We have made use of NASA data products, including WISE and NED,
and also NSF OIR Lab data products. We acknowledge the use of the HyperLeda database.
HC is grateful to the Institut Universitaire de France and CNES for its support. KS, CH, and TD are supported by the Australian Government through the Australian Research Council's Laureate Fellowship funding scheme (project FL180100168). AD acknowledges financial support from the Project IDEXLYON at the University of Lyon under the Investments for the Future Program (ANR-16-IDEX-0005). AB acknowledges support from the Centre National d'Etudes Spatiales (CNES), France. DK acknowledges funding from the European Research Council (ERC) under the European Union’s Horizon 2020
research and innovation programme (grant agreement no. 679627; project name FORNAX. RCKK is supported by the South African Research Chairs Initiative of the Department of Science and Technology
and National Research Foundation. FB acknowledges funding from the European Research Council (ERC) under the European Union’s Horizon 2020 research and innovation programme (grant agreement No.726384/Empire). AD is supported by a KIAS Individual Grant (PG087201) at Korea Institute for Advanced Study. Many thanks to Dmitry Makarov for helping us on the cross-identification with galaxy PGC names.

%%%%%%%%%%%%%%%%%%%%%%%%%%%%%%%%%%%%%%%%%%%%%%%%%%
%\section*{Data Availability}
%Distances, HI linewidths, and WISE magnitudes are available for the four fields in the online supplementary material. All other data from the parent catalogues will be shared on reasonable request to the corresponding author.



\bibliographystyle{mnras}
\bibliography{references} % if your bibtex file is called example.bib
\clearpage

 \appendix

\section{DATA}
%The 21cm HI, WISE photometery, and TF distances catalog and plots of the HI spectra are available electronically. 
Tables in Appendix \ref{data} present four catalogs of Hydra, Norma, Eridanus, and NGC4636. The parameters listed in the catalogs are:\\

Column (1) - WALLABY ID as reported in the WALLABY Pilot Survey data release (Westmeier et al., in prep.).

Columns (2) - Principal Galaxies Catalogue number (PGC)

Columns (3 and 4) - Right Ascension (RA) and Declination (Dec.) in the J2000.0 epoch of the fitted position in WALLABY.

Column (5) - WISE W1 total magnitude after Galactic extinction and k corrections.

Column (6) - Inclination calculated from WISE ($b/a$) as in equation \ref{inclination}.

Column (7) - Linewidths corrected for inclination.

Column (8) - CMB redshifts.

Column (9) - Tully-Fisher luminosity distances in Mpc.

Column (10 and 11) - Distance modulus and associated error.\\

\onecolumn
\label{data}
\clearpage
\begin{longtable}{lrrrrrrrrrr}
\caption[]{Hydra HI data from WALLABY phase 1 pilot survey.}\\
\hline\hline
  \multicolumn{1}{c}{Name} &
  \multicolumn{1}{c}{PGC} &
  \multicolumn{1}{c}{RA} &
  \multicolumn{1}{c}{DEC} &
  \multicolumn{1}{c}{$W1_{tot}$} &
  \multicolumn{1}{c}{$i$} &
  \multicolumn{1}{c}{$W_{mx}$} &
  \multicolumn{1}{c}{$z_{cmb}$} &
  \multicolumn{1}{c}{$d_L$} &
  \multicolumn{1}{c}{$\mu$} &
  \multicolumn{1}{c}{$e_{\mu}$} \\
\hline
  J100342-270137 & 29166 & 150.9256 & -27.0271 & 10.715 & 79.3316 & 185.2012 & 0.00430152 & 20.0 & 31.5 & 0.58\\
  J100351-273417 & 29179 & 150.9662 & -27.5715 & 8.947 & 51.0182 & 421.9486 & 0.01043229 & 42.1 & 33.12 & 0.58\\
  J100426-282638 & 29216 & 151.1091 & -28.4442 & 9.9561 & 73.1772 & 202.6735 & 0.00471975 & 16.7 & 31.12 & 0.58\\
  J100640-273917 & 29378 & 151.667 & -27.6549 & 12.923 & 62.8353 & 223.6714 & 0.01536106 & 79.0 & 34.49 & 0.58\\
  J100808-260942 & 29470 & 152.0353 & -26.1618 & 13.4138 & 57.4373 & 132.8899 & 0.01141974 & 36.9 & 32.84 & 0.58\\
  J100830-262140 & 768685 & 152.1256 & -26.3611 & 14.5217 & 76.0108 & 118.515 & 0.01512221 & 49.5 & 33.47 & 0.58\\
  J100939-290112 & 154738 & 152.4128 & -29.0202 & 12.4737 & 47.9538 & 278.7484 & 0.0480433 & 97.4 & 34.94 & 0.58\\
  J101025-275214 & 748738 & 152.6054 & -27.8708 & 13.1662 & 68.9346 & 132.8803 & 0.00919987 & 33.0 & 32.59 & 0.58\\
  J101208-252000 & 154861 & 153.0362 & -25.3335 & 12.2087 & 58.4722 & 305.026 & 0.04684504 & 102.3 & 35.05 & 0.58\\
  J101221-254604 & 29719 & 153.0904 & -25.7678 & 11.4898 & 60.6661 & 430.1545 & 0.03330433 & 140.9 & 35.74 & 0.58\\
  J101247-275028 & 29743 & 153.1998 & -27.8413 & 9.4937 & 76.0108 & 350.3921 & 0.00984753 & 38.1 & 32.9 & 0.58\\
  J101247-291053 & 732369 & 153.1962 & -29.1815 & 13.2847 & 61.97 & 267.361 & 0.03197037 & 130.8 & 35.58 & 0.58\\
  J101348-273147 & 29821 & 153.4516 & -27.5299 & 13.3432 & 57.7338 & 164.3848 & 0.00981759 & 53.5 & 33.64 & 0.58\\
  J101357-262550 &  & 153.4904 & -26.4308 & 15.1772 & 52.8182 & 102.9217 & 0.03382259 & 51.3 & 33.55 & 0.58\\
  J101359-253824 & 29836 & 153.4964 & -25.64 & 13.2213 & 71.1098 & 205.0435 & 0.01177462 & 76.9 & 34.43 & 0.58\\
  J101359-274538 &  & 153.4978 & -27.7607 & 15.9193 & 73.7025 & 70.8468 & 0.00969587 & 35.6 & 32.76 & 0.59\\
  J101438-272431 & 29888 & 153.661 & -27.4087 & 10.8963 & 54.1261 & 386.2726 & 0.01496544 & 87.4 & 34.71 & 0.58\\
  J101441-285221 & 29892 & 153.6744 & -28.8726 & 7.4814 & 71.9901 & 312.3018 & 0.00479612 & 12.1 & 30.42 & 0.58\\
  J101448-285723 & 29903 & 153.7006 & -28.9565 & 10.5448 & 72.9533 & 400.5999 & 0.01509413 & 79.7 & 34.51 & 0.58\\
  J101526-264259 & 764730 & 153.8614 & -26.7166 & 13.8652 & 47.6238 & 300.5138 & 0.03412958 & 213.3 & 36.64 & 0.58\\
  J101531-292423 & 29945 & 153.8793 & -29.4066 & 12.4099 & 66.2786 & 337.516 & 0.03173132 & 135.9 & 35.67 & 0.58\\
  J101537-272427 &  & 153.9057 & -27.4078 & 15.3554 & 49.2591 & 96.3481 & 0.01157047 & 49.1 & 33.46 & 0.58\\
  J101632-291258 &  & 154.1373 & -29.2163 & 13.0964 & 76.0108 & 176.2266 & 0.01514422 & 54.5 & 33.68 & 0.64\\
  J101756-285557 &  & 154.4866 & -28.9325 & 14.4099 & 67.8558 & 170.5825 & 0.03769535 & 93.8 & 34.86 & 0.58\\
  J101834-281550 & 3753792 & 154.6456 & -28.264 & 14.6891 & 47.6238 & 151.6106 & 0.04194573 & 85.3 & 34.65 & 0.59\\
  J101920-264135 & 30161 & 154.8353 & -26.6931 & 12.3029 & 56.6184 & 318.5536 & 0.04414246 & 116.0 & 35.32 & 0.58\\
  J101927-264159 & 764945 & 154.8662 & -26.6999 & 14.1879 & 50.3033 & 110.4705 & 0.00964968 & 37.2 & 32.85 & 0.58\\
  J102017-253913 & 30216 & 155.0743 & -25.6538 & 13.4071 & 51.6485 & 150.4681 & 0.01264661 & 46.6 & 33.34 & 0.58\\
  J102019-285220 & 30218 & 155.0828 & -28.8722 & 11.15 & 76.9694 & 429.0481 & 0.03237219 & 119.9 & 35.39 & 0.58\\
  J102023-253050 &  & 155.0966 & -25.514 & 15.3675 & 58.4722 & 132.569 & 0.01211333 & 90.4 & 34.78 & 0.58\\
  J102030-260951 & 30234 & 155.1281 & -26.1642 & 12.8541 & 48.6094 & 417.2113 & 0.0572516 & 249.2 & 36.98 & 0.58\\
  J102100-273339 & 753130 & 155.2535 & -27.5608 & 14.1518 & 55.7186 & 254.1508 & 0.03442685 & 177.2 & 36.24 & 0.58\\
  J102115-250342 & 783072 & 155.3137 & -25.0618 & 13.6738 & 52.5855 & 313.4991 & 0.03393395 & 211.6 & 36.63 & 0.58\\
  J102233-294017 & 726509 & 155.6378 & -29.6715 & 14.9756 & 47.6238 & 200.3425 & 0.02718859 & 165.0 & 36.09 & 0.58\\
  J102328-253424 & 777390 & 155.8685 & -25.5736 & 12.4153 & 68.5745 & 399.6161 & 0.04170041 & 187.7 & 36.37 & 0.58\\
  J102338-264531 & 764220 & 155.9116 & -26.7589 & 14.1452 & 53.9731 & 290.575 & 0.03581803 & 227.6 & 36.79 & 0.58\\
  J102416-270241 & 760541 & 156.0683 & -27.0448 & 14.424 & 47.4581 & 261.9494 & 0.03578983 & 212.7 & 36.64 & 0.58\\
  J102430-290904 & 732726 & 156.1259 & -29.1511 & 13.6075 & 54.5075 & 143.7009 & 0.01363351 & 46.8 & 33.35 & 0.58\\
  J102439-274841 & 749549 & 156.1626 & -27.8117 & 15.1621 & 49.5819 & 136.6024 & 0.01315652 & 87.1 & 34.7 & 0.58\\
  J102600-280334 &  & 156.5018 & -28.0597 & 15.2205 & 54.9633 & 173.4279 & 0.0375637 & 140.5 & 35.74 & 0.58\\
  J102621-291150 & 732211 & 156.5886 & -29.1973 & 13.6973 & 62.403 & 97.0405 & 0.01345127 & 23.2 & 31.83 & 0.58\\
  J102637-264142 & 764986 & 156.6545 & -26.6952 & 14.5557 & 57.7338 & 112.3494 & 0.0325455 & 45.5 & 33.29 & 0.59\\
  J102643-261256 & 770222 & 156.6797 & -26.2157 & 12.7449 & 51.6485 & 345.5666 & 0.03439577 & 165.9 & 36.1 & 0.58\\
  J102747-283425 &  & 156.9461 & -28.5738 & 15.4282 & 48.6909 & 238.2983 & 0.03272924 & 282.3 & 37.25 & 0.58\\
  J102818-255446 & 773700 & 157.0764 & -25.9129 & 16.0183 & 52.43 & 93.3625 & 0.01386497 & 62.8 & 33.99 & 0.58\\
  J102934-261937 & 30912 & 157.3949 & -26.3271 & 12.8164 & 47.2921 & 217.74 & 0.01427068 & 71.5 & 34.27 & 0.58\\
  J103002-284116 & 738653 & 157.5091 & -28.688 & 13.1147 & 61.3915 & 124.1583 & 0.01394284 & 28.3 & 32.26 & 0.58\\
  J103015-270743 & 759425 & 157.5651 & -27.1287 & 14.5215 & 69.0066 & 215.2906 & 0.03749136 & 153.4 & 35.93 & 0.58\\
  J103114-295837 & 722930 & 157.809 & -29.9771 & 14.6759 & 59.2068 & 109.4269 & 0.01511548 & 45.7 & 33.3 & 0.58\\
  J103139-273049 & 753808 & 157.9143 & -27.5138 & 14.3307 & 51.334 & 165.215 & 0.01305822 & 85.1 & 34.65 & 0.58\\
  J103141-300815 & 720959 & 157.921 & -30.1377 & 14.8566 & 62.8353 & 97.786 & 0.01849056 & 40.2 & 33.02 & 0.58\\
  J103240-282058 & 742765 & 158.169 & -28.3495 & 12.6848 & 71.4757 & 181.3984 & 0.01422809 & 47.6 & 33.39 & 0.58\\
  J103241-273137 & 31149 & 158.1721 & -27.5271 & 12.0583 & 71.4757 & 147.6499 & 0.01391053 & 24.2 & 31.91 & 0.58\\
  J103257-250937 & 782023 & 158.2381 & -25.1609 & 14.5155 & 62.1144 & 243.2447 & 0.02565921 & 192.8 & 36.43 & 0.62\\
  J103258-274013 & 31171 & 158.2426 & -27.6711 & 13.0522 & 70.8908 & 126.9981 & 0.01160875 & 28.7 & 32.29 & 0.58\\
  J103348-271429 & 757813 & 158.452 & -27.2417 & 13.1222 & 56.1694 & 346.7011 & 0.03655502 & 198.6 & 36.49 & 0.58\\
  J103353-274945 & 31238 & 158.4748 & -27.8291 & 12.2977 & 60.448 & 214.9652 & 0.01032262 & 54.9 & 33.7 & 0.58\\
  J103359-301003 & 31242 & 158.4966 & -30.1677 & 9.8673 & 78.7202 & 407.8787 & 0.01277512 & 60.3 & 33.9 & 0.58\\
  J103420-265408 & 92288 & 158.5865 & -26.9023 & 14.011 & 78.9802 & 135.4984 & 0.01362264 & 50.5 & 33.51 & 0.58\\
  J103502-293019 & 728536 & 158.7604 & -29.5062 & 12.3451 & 49.8231 & 111.2483 & 0.01362627 & 16.1 & 31.04 & 0.58\\
  J103507-275923 & 31334 & 158.783 & -27.9905 & 12.1861 & 73.7025 & 168.7821 & 0.00926373 & 33.0 & 32.59 & 0.58\\
  J103521-274137 & 31355 & 158.8391 & -27.6938 & 13.2414 & 75.6965 & 175.4385 & 0.01079559 & 57.7 & 33.81 & 0.58\\
  J103547-290131 & 31390 & 158.9463 & -29.0252 & 13.0973 & 50.2235 & 312.2781 & 0.03381084 & 161.0 & 36.03 & 0.58\\
  J103609-244856 & 31415 & 159.0412 & -24.8154 & 13.0609 & 53.8966 & 84.1631 & 0.00454496 & 13.2 & 30.61 & 0.58\\
  J103645-281010 & 744989 & 159.1902 & -28.1698 & 13.2636 & 49.0161 & 162.9367 & 0.01278334 & 50.7 & 33.53 & 0.58\\
  J103646-293253 & 31484 & 159.196 & -29.5482 & 13.2256 & 51.334 & 197.2334 & 0.01304613 & 71.5 & 34.27 & 0.58\\
  J103650-270902 & 31490 & 159.2096 & -27.1485 & 13.177 & 67.8558 & 292.5814 & 0.03859519 & 147.7 & 35.85 & 0.58\\
  J103651-260227 & 31491 & 159.2131 & -26.0412 & 13.526 & 49.8231 & 204.1734 & 0.00907223 & 87.7 & 34.72 & 0.58\\
  J103653-270311 & 31494 & 159.2211 & -27.0532 & 12.0485 & 54.6596 & 234.1461 & 0.01299891 & 57.6 & 33.8 & 0.58\\
  J103655-265412 & 762452 & 159.233 & -26.9029 & 13.0441 & 64.2723 & 199.8074 & 0.01197818 & 67.4 & 34.14 & 0.58\\
  J103701-284018 & 738882 & 159.2551 & -28.6725 & 14.2125 & 65.4907 & 236.2912 & 0.03368768 & 158.7 & 36.0 & 0.58\\
  J103704-252038 & 31514 & 159.2683 & -25.3441 & 12.3722 & 76.8078 & 199.2584 & 0.01353874 & 49.2 & 33.46 & 0.58\\
  J103737-261641 & 31574 & 159.4074 & -26.2776 & 10.7246 & 53.0503 & 322.8375 & 0.01385994 & 57.5 & 33.8 & 0.58\\
  J103803-281007 & 31602 & 159.5132 & -28.1688 & 14.1108 & 72.5815 & 304.9854 & 0.03634067 & 245.6 & 36.95 & 0.58\\
  J103809-260453 & 31612 & 159.5414 & -26.0815 & 13.0599 & 72.73 & 144.5152 & 0.0130769 & 36.8 & 32.83 & 0.58\\
  J103818-285307 & 31626 & 159.5743 & -28.8845 & 11.4949 & 74.69 & 297.56 & 0.01578427 & 70.3 & 34.23 & 0.58\\
  J103828-283056 & 740766 & 159.6196 & -28.5157 & 13.8722 & 52.6631 & 169.7936 & 0.01568042 & 72.6 & 34.3 & 0.58\\
  J103840-283405 & 31642 & 159.6679 & -28.5684 & 10.4866 & 64.4158 & 323.7424 & 0.01272478 & 51.8 & 33.57 & 0.58\\
  J103858-300500 & 3758554 & 159.7422 & -30.084 & 15.1371 & 63.8418 & 81.3298 & 0.01370769 & 32.2 & 32.54 & 0.6\\
  J103902-291255 & 31664 & 159.7607 & -29.2144 & 13.9916 & 62.6913 & 165.4385 & 0.01181481 & 73.0 & 34.32 & 0.58\\
  J103905-265519 & 93061 & 159.7724 & -26.9228 & 12.8866 & 66.2069 & 350.8164 & 0.0363276 & 182.2 & 36.3 & 0.58\\
  J103918-265030 & 31683 & 159.8261 & -26.8421 & 10.3241 & 72.1376 & 344.6117 & 0.01142924 & 54.1 & 33.67 & 0.58\\
  J103922-293505 & 31690 & 159.8422 & -29.5846 & 11.6635 & 48.8537 & 274.8887 & 0.01408946 & 65.4 & 34.08 & 0.58\\
  J104000-292445 & 31732 & 160.0014 & -29.4123 & 13.127 & 66.3502 & 184.4949 & 0.01371067 & 60.3 & 33.9 & 0.58\\
  J104004-301606 & 31738 & 160.0178 & -30.2679 & 10.59 & 65.634 & 286.5209 & 0.01248467 & 43.1 & 33.17 & 0.58\\
  J104058-274546 & 750199 & 160.2427 & -27.7623 & 14.349 & 67.1384 & 162.7876 & 0.01446573 & 83.4 & 34.61 & 0.58\\
  J104059-270456 & 31805 & 160.2475 & -27.0819 & 9.6466 & 56.2443 & 396.914 & 0.01698876 & 51.8 & 33.57 & 0.58\\
  J104100-284430 & 31809 & 160.2552 & -28.7418 & 12.2681 & 55.5679 & 192.7746 & 0.01365658 & 44.1 & 33.22 & 0.58\\
  J104139-254049 &  & 160.4145 & -25.6805 & 14.4528 & 54.6596 & 122.5896 & 0.01404383 & 51.2 & 33.54 & 0.58\\
  J104139-274639 & 31852 & 160.4136 & -27.778 & 13.753 & 59.865 & 136.4406 & 0.01569824 & 45.4 & 33.28 & 0.58\\
  J104157-264302 &  & 160.4888 & -26.7174 & 13.6008 & 63.4108 & 256.0838 & 0.03786389 & 139.5 & 35.72 & 0.58\\
  J104239-300357 &  & 160.6646 & -30.0664 & 15.6547 & 69.6565 & 82.1224 & 0.00940569 & 41.7 & 33.1 & 0.6\\
  J104245-264738 & 31924 & 160.6913 & -26.7937 & 12.522 & 57.7338 & 345.3264 & 0.03673843 & 149.5 & 35.87 & 0.58\\
  J104252-252014 &  & 160.7192 & -25.3373 & 13.8052 & 48.935 & 126.0006 & 0.02245921 & 40.0 & 33.01 & 0.58\\
  J104311-261500 & 31951 & 160.7974 & -26.2501 & 10.1873 & 61.1016 & 430.6217 & 0.01629157 & 77.5 & 34.45 & 0.58\\
  J104326-251857 & 31960 & 160.859 & -25.316 & 13.5057 & 48.6094 & 125.2967 & 0.01373015 & 34.5 & 32.69 & 0.58\\
  J104339-285157 & 31981 & 160.913 & -28.8661 & 10.9693 & 76.7273 & 283.5747 & 0.01271548 & 50.4 & 33.51 & 0.58\\
  J104414-271548 & 757564 & 161.0589 & -27.2637 & 14.1818 & 59.1335 & 226.0111 & 0.03779767 & 143.8 & 35.79 & 0.58\\
  J104442-290119 & 734399 & 161.1761 & -29.0214 & 15.2513 & 59.1335 & 137.4706 & 0.03477507 & 91.8 & 34.81 & 0.58\\
  J104524-251723 & 780619 & 161.3536 & -25.2898 & 13.9341 & 50.9391 & 271.7404 & 0.04946774 & 181.9 & 36.3 & 0.58\\
  J104620-293733 & 727073 & 161.586 & -29.626 & 13.0861 & 56.9913 & 269.5008 & 0.04061149 & 121.2 & 35.42 & 0.58\\
  J104629-253308 & 32175 & 161.6229 & -25.5523 & 14.8472 & 65.7056 & 145.9224 & 0.01318261 & 85.3 & 34.66 & 0.59\\
\hline\end{longtable}

\clearpage
\begin{longtable}{lrrrrrrrrrr}
\caption[]{Norma HI data from WALLABY phase 1 pilot survey.}\\
\hline\hline
  \multicolumn{1}{c}{Name} &
  \multicolumn{1}{c}{PGC} &
  \multicolumn{1}{c}{RA} &
  \multicolumn{1}{c}{DEC} &
  \multicolumn{1}{c}{$W1_{tot}$} &
  \multicolumn{1}{c}{$i$} &
  \multicolumn{1}{c}{$W_{mx}$} &
  \multicolumn{1}{c}{$z_{cmb}$} &
  \multicolumn{1}{c}{$d_L$} &
  \multicolumn{1}{c}{$\mu$} &
  \multicolumn{1}{c}{$e_{\mu}$} \\
\hline
  J163435-620248 & 58527 & 248.647418 & -62.046778 & 12.5906 & 62.8353 & 234.9112 & 0.01511102 & 74.4 & 34.36 & 0.58\\
  J163749-621352 & 4077297 & 249.454206 & -62.231162 & 10.5802 & 54.7356 & 175.1385 & 0.01625367 & 16.9 & 31.14 & 0.58\\
  J163927-594844 & 3078607 & 249.866523 & -59.812295 & 12.0403 & 67.2818 & 471.588 & 0.03785902 & 216.1 & 36.67 & 0.58\\
  J164107-605902 & 58755 & 250.281497 & -60.983986 & 13.5023 & 80.7086 & 160.1005 & 0.01186631 & 54.7 & 33.69 & 0.59\\
  J164113-603202 &  & 250.307194 & -60.534109 & 13.7705 & 56.9913 & 186.0271 & 0.03986854 & 82.3 & 34.58 & 0.58\\
  J164206-613441 & 3078658 & 250.528374 & -61.578262 & 14.1884 & 71.4757 & 191.9448 & 0.01770498 & 105.9 & 35.12 & 0.58\\
  J164249-610527 & 92488 & 250.707724 & -61.091025 & 10.6964 & 79.8702 & 310.8455 & 0.01506462 & 52.8 & 33.61 & 0.58\\
  J164355-620233 & 349886 & 250.981502 & -62.042582 & 12.6012 & 52.973 & 379.5319 & 0.03933249 & 185.4 & 36.34 & 0.58\\
  J164437-605103 & 58893 & 251.156286 & -60.851094 & 10.5273 & 72.0638 & 424.6372 & 0.01781932 & 88.3 & 34.73 & 0.58\\
  J164455-575030 & 3078727 & 251.230899 & -57.841714 & 12.282 & 47.209 & 400.6343 & 0.04826815 & 177.3 & 36.24 & 0.58\\
  J164508-611614 &  & 251.285273 & -61.27071 & 13.1157 & 79.3316 & 171.9725 & 0.01246623 & 52.5 & 33.6 & 0.58\\
  J164727-575143 & 3078771 & 251.863957 & -57.862209 & 13.2117 & 62.8353 & 361.9206 & 0.04771013 & 224.5 & 36.76 & 0.58\\
  J164739-570817 & 393557 & 251.914425 & -57.138085 & 14.3561 & 54.7356 & 83.2827 & 0.00287685 & 23.5 & 31.86 & 0.59\\
  J164749-613936 &  & 251.954282 & -61.660226 & 12.805 & 63.5545 & 174.2319 & 0.01770366 & 46.6 & 33.34 & 0.58\\
  J164826-623040 & 343635 & 252.108723 & -62.511225 & 11.9037 & 62.8353 & 185.4562 & 0.01496239 & 34.6 & 32.7 & 0.58\\
  J164911-620001 & 350325 & 252.298069 & -62.000386 & 12.6063 & 57.7338 & 275.5516 & 0.05177993 & 101.3 & 35.03 & 0.58\\
  J165131-603514 & 59130 & 252.882699 & -60.587289 & 12.5929 & 64.9176 & 133.5985 & 0.01132893 & 25.6 & 32.04 & 0.58\\
  J165439-605730 &  & 253.663341 & -60.95855 & 14.0336 & 69.2952 & 168.9091 & 0.01054331 & 77.4 & 34.44 & 0.58\\
  J165455-610856 & 357773 & 253.730344 & -61.149002 & 15.647 & 54.7356 & 105.3281 & 0.01482964 & 66.5 & 34.11 & 0.59\\
  J165520-592615 & 3078940 & 253.834941 & -59.437529 & 13.5311 & 75.2293 & 153.058 & 0.01655333 & 50.9 & 33.54 & 0.58\\
  J165559-592439 & 165948 & 253.997339 & -59.410888 & 11.7775 & 47.6238 & 265.3185 & 0.01615457 & 64.4 & 34.04 & 0.58\\
  J165632-594959 & 4584387 & 254.134589 & -59.833236 & 11.8378 & 54.5836 & 468.7337 & 0.04802051 & 194.6 & 36.45 & 0.58\\
  J165644-622413 & 4075676 & 254.184255 & -62.403763 & 9.9351 & 64.3441 & 377.1867 & 0.01719366 & 53.7 & 33.65 & 0.58\\
  J165645-620555 &  & 254.191411 & -62.098772 & 15.3289 & 58.4722 & 195.9205 & 0.03011796 & 186.1 & 36.35 & 0.59\\
  J165840-582909 & 141869 & 254.669097 & -58.485927 & 11.0622 & 54.7356 & 307.411 & 0.02063727 & 61.2 & 33.93 & 0.58\\
  J165847-612501 &  & 254.69821 & -61.416959 & 14.2337 & 74.4606 & 160.8808 & 0.01597719 & 77.4 & 34.44 & 0.59\\
  J165934-623256 & 4080474 & 254.894684 & -62.548998 & 10.5454 & 57.1401 & 255.9524 & 0.01642583 & 34.1 & 32.66 & 0.58\\
  J170039-584218 & 377019 & 255.16626 & -58.705246 & 13.4896 & 69.2952 & 183.8757 & 0.02094197 & 70.8 & 34.25 & 0.59\\
  J170116-621215 & 165971 & 255.317106 & -62.204321 & 12.073 & 55.039 & 406.3242 & 0.04637359 & 165.4 & 36.09 & 0.58\\
  J170149-603901 & 165952 & 255.455525 & -60.650362 & 12.2495 & 65.3475 & 290.4757 & 0.01649612 & 95.0 & 34.89 & 0.58\\
  J170232-584322 & 165953 & 255.634453 & -58.722864 & 11.5633 & 60.1568 & 314.7373 & 0.02064314 & 80.6 & 34.53 & 0.58\\
  J170349-595038 &  & 255.954693 & -59.844105 & 13.0267 & 49.0161 & 253.0155 & 0.01982964 & 104.6 & 35.1 & 1.1\\
  J170700-600654 & 165955 & 256.751847 & -60.11503 & 12.2699 & 62.1144 & 227.4055 & 0.01436001 & 60.3 & 33.9 & 0.58\\
  J170747-595059 & 59659 & 256.947861 & -59.849852 & 11.7537 & 79.1552 & 281.0189 & 0.01579552 & 71.0 & 34.26 & 0.58\\
  J171120-603454 & 362143 & 257.833867 & -60.581908 & 14.2512 & 50.0636 & 95.2061 & 0.01670085 & 28.9 & 32.3 & 0.58\\
  J171134-610802 &  & 257.893397 & -61.134105 & 16.0695 & 54.7356 & 97.9796 & 0.00902464 & 70.5 & 34.24 & 0.6\\
  J171309-603124 & 362690 & 258.287752 & -60.523391 & 13.4709 & 70.4539 & 125.2158 & 0.01621556 & 33.9 & 32.65 & 0.58\\
  J171339-615822 & 92494 & 258.415875 & -61.973005 & 15.4747 & 64.2723 & 152.0756 & 0.01505209 & 123.2 & 35.45 & 0.59\\
  J171538-602409 &  & 258.909689 & -60.402582 & 13.0023 & 79.9614 & 243.7314 & 0.0195147 & 96.4 & 34.92 & 0.58\\
  J171558-590923 &  & 258.994367 & -59.156581 & 13.0923 & 72.0638 & 189.1948 & 0.01186545 & 62.2 & 33.97 & 0.58\\
  J171850-601104 & 365445 & 259.711772 & -60.184587 & 13.7175 & 47.1257 & 253.8042 & 0.01551251 & 144.7 & 35.8 & 0.58\\
  J171924-615016 & 92496 & 259.853206 & -61.837997 & 13.265 & 75.2293 & 276.1248 & 0.02455815 & 137.8 & 35.7 & 0.58\\
\hline\end{longtable}

\clearpage
\begin{longtable}{lrrrrrrrrrr}
\caption[]{Eridanus HI data from WALLABY phase 1 pilot survey.}\\
\hline\hline
  \multicolumn{1}{c}{Name} &
  \multicolumn{1}{c}{PGC} &
  \multicolumn{1}{c}{RA} &
  \multicolumn{1}{c}{DEC} &
  \multicolumn{1}{c}{$W1_{tot}$} &
  \multicolumn{1}{c}{$i$} &
  \multicolumn{1}{c}{$W_{mx}$} &
  \multicolumn{1}{c}{$z_{cmb}$} &
  \multicolumn{1}{c}{$d_L$} &
  \multicolumn{1}{c}{$\mu$} &
  \multicolumn{1}{c}{$e_{\mu}$} \\
\hline
  J034114-235017 & 13561 & 55.308333 & -23.838056 & 10.1562 & 51.4127 & 254.5868 & 0.00591068 & 28.2 & 32.25 & 0.58\\
  J033859-223502 & 13442 & 54.745833 & -22.583889 & 13.3234 & 62.5472 & 171.2887 & 0.01408583 & 57.3 & 33.79 & 0.58\\
  J034056-223350 & 13544 & 55.233333 & -22.563889 & 8.0092 & 64.7743 & 374.7365 & 0.00478841 & 21.8 & 31.7 & 0.58\\
  J034517-230001 & 13760 & 56.320833 & -23.000278 & 9.306 & 56.1694 & 361.1469 & 0.00483078 & 37.0 & 32.84 & 0.58\\
  J034219-224520 & 13608 & 55.579167 & -22.755556 & 11.9271 & 51.6485 & 145.3675 & 0.00485579 & 22.1 & 31.72 & 0.58\\
  J033941-235054 & 13477 & 54.920833 & -23.848333 & 13.8833 & 82.0066 & 71.6966 & 0.00503014 & 14.2 & 30.77 & 0.58\\
  J034057-214245 & 13543 & 55.2375 & -21.7125 & 11.5644 & 74.9975 & 148.0462 & 0.00527728 & 19.3 & 31.43 & 0.58\\
  J034040-221711 & 13531 & 55.166667 & -22.286389 & 12.4245 & 77.6238 & 154.5925 & 0.00553255 & 31.2 & 32.47 & 0.58\\
  J034002-192200 & 13491 & 55.008333 & -19.366667 & 12.0881 & 80.6134 & 114.5336 & 0.00365656 & 15.1 & 30.9 & 0.58\\
  J034814-212824 & 13871 & 57.058333 & -21.473333 & 9.6576 & 55.1904 & 274.0381 & 0.00493632 & 25.8 & 32.06 & 0.58\\
  J034337-211418 & 13689 & 55.904167 & -21.238333 & 12.4218 & 72.5815 & 128.9114 & 0.00499355 & 22.1 & 31.72 & 0.58\\
  J034434-211123 & 135119 & 56.141667 & -21.189722 & 13.1936 & 70.0185 & 72.3556 & 0.00488434 & 10.6 & 30.12 & 0.58\\
  J034131-214051 & 13569 & 55.379167 & -21.680833 & 10.3246 & 74.4606 & 157.7669 & 0.00506959 & 12.3 & 30.45 & 0.58\\
  J033921-212450 & 13460 & 54.8375 & -21.413889 & 11.9631 & 66.4218 & 117.8377 & 0.00503156 & 15.1 & 30.89 & 0.58\\
  J033537-211742 & 13283 & 53.904167 & -21.295 & 11.8244 & 75.2293 & 146.8529 & 0.00562052 & 21.5 & 31.66 & 0.58\\
  J033341-212844 & 13184 & 53.420833 & -21.478889 & 8.9854 & 48.6094 & 265.2557 & 0.00578946 & 17.8 & 31.25 & 0.58\\
  J033347-192946 & 13190 & 53.445833 & -19.496111 & 10.187 & 54.7356 & 231.4768 & 0.00615231 & 23.9 & 31.89 & 0.58\\
  J033810-194412 & 13401 & 54.541667 & -19.736667 & 12.6368 & 71.7693 & 175.8256 & 0.01352603 & 43.9 & 33.21 & 0.58\\
  J033757-213938 & 13392 & 54.4875 & -21.660556 & 13.5581 & 73.552 & 136.5896 & 0.01347565 & 41.6 & 33.09 & 0.58\\
  J033147-211309 & 831121 & 52.945833 & -21.219167 & 14.354 & 64.9892 & 101.5197 & 0.01371815 & 34.2 & 32.67 & 0.58\\
  J032455-214701 & 12762 & 51.229167 & -21.783611 & 12.8097 & 55.1904 & 102.3076 & 0.00439474 & 17.0 & 31.16 & 0.58\\
  J032425-213233 & 12737 & 51.104167 & -21.5425 & 8.4969 & 72.5815 & 341.6675 & 0.00484116 & 23.0 & 31.8 & 0.58\\
  J033527-211302 & 831168 & 53.8625 & -21.217222 & 12.8515 & 48.4461 & 120.2676 & 0.00466241 & 23.6 & 31.86 & 0.58\\
  J032735-211339 & 12889 & 51.895833 & -21.2275 & 11.3398 & 75.7749 & 166.0928 & 0.00517694 & 21.7 & 31.68 & 0.58\\
  J032831-222957 & 813307 & 52.129167 & -22.499167 & 13.6391 & 73.7025 & 69.8049 & 0.00548873 & 12.1 & 30.41 & 0.58\\
  J033653-245445 & 13342 & 54.220833 & -24.9125 & 12.2326 & 58.3248 & 124.5537 & 0.00575066 & 19.0 & 31.39 & 0.58\\
  J033302-240756 & 13154 & 53.258333 & -24.132222 & 12.6236 & 71.4757 & 162.4149 & 0.00598226 & 37.5 & 32.87 & 0.58\\
  J033617-253615 & 13304 & 54.070833 & -25.604167 & 11.4992 & 67.3535 & 124.6075 & 0.00493199 & 13.5 & 30.66 & 0.58\\
  J032937-232103 & 12986 & 52.404167 & -23.350833 & 12.9769 & 74.3844 & 116.2925 & 0.00509535 & 23.5 & 31.85 & 0.58\\
  J033326-234246 & 13171 & 53.358333 & -23.712778 & 9.5617 & 76.0108 & 256.6107 & 0.00564806 & 21.8 & 31.69 & 0.58\\
  J033228-232245 & 13127 & 53.116667 & -23.379167 & 13.3847 & 56.3192 & 69.6998 & 0.00543831 & 10.7 & 30.15 & 0.58\\
\hline\end{longtable}

\clearpage
\begin{longtable}{lrrrrrrrrrr}
\caption[]{NGC4636 HI data from WALLABY phase 1 pilot survey.}\\
\hline\hline
  \multicolumn{1}{c}{Name} &
  \multicolumn{1}{c}{PGC} &
  \multicolumn{1}{c}{RA} &
  \multicolumn{1}{c}{DEC} &
  \multicolumn{1}{c}{$W1_{tot}$} &
  \multicolumn{1}{c}{$i$} &
  \multicolumn{1}{c}{$W_{mx}$} &
  \multicolumn{1}{c}{$z_{cmb}$} &
  \multicolumn{1}{c}{$d_L$} &
  \multicolumn{1}{c}{$\mu$} &
  \multicolumn{1}{c}{$e_{\mu}$} \\
\hline
  J122708+055255 & 40801 & 186.78688 & 5.88209 & 10.7665 & 80.2379 & 166.4096 & 0.00485391 & 16.7 & 31.12 & 0.58\\
  J122713-020236 & 40820 & 186.8062 & -2.04356 & 11.9998 & 64.9892 & 82.7606 & 0.02268628 & 7.9 & 29.48 & 0.58\\
  J122729+073841 & 40861 & 186.87277 & 7.64494 & 12.0952 & 58.1773 & 141.229 & 0.00401966 & 22.6 & 31.77 & 0.58\\
  J122745+013601 & 135803 & 186.93906 & 1.60045 & 15.007 & 48.4461 & 66.8153 & 0.00548086 & 20.9 & 31.6 & 0.6\\
  J122755+054312 & 40933 & 186.9813 & 5.72019 & 11.3223 & 77.0504 & 138.523 & 0.00864351 & 15.3 & 30.92 & 0.58\\
  J122852+041739 & 41088 & 187.21861 & 4.2942 & 11.4152 & 81.1938 & 289.4117 & 0.01528698 & 64.3 & 34.04 & 0.58\\
  J122930+074142 & 41170 & 187.37658 & 7.69522 & 10.983 & 67.2818 & 190.8034 & 0.00363512 & 23.9 & 31.89 & 0.58\\
  J122932+005020 & 41177 & 187.38499 & 0.83914 & 14.2449 & 74.537 & 121.3941 & 0.00862244 & 45.6 & 33.3 & 0.58\\
  J123021-013813 & 3294304 & 187.59016 & -1.63716 & 15.4059 & 59.5728 & 73.0627 & 0.00827984 & 29.8 & 32.37 & 0.59\\
  J123022+034431 & 41307 & 187.59204 & 3.74218 & 10.8261 & 77.6238 & 317.3754 & 0.01525979 & 58.3 & 33.83 & 0.58\\
  J123026+041450 & 41317 & 187.61136 & 4.24724 & 9.5804 & 53.5128 & 385.5769 & 0.01525759 & 47.5 & 33.38 & 0.58\\
  J123132+071828 & 5060276 & 187.88433 & 7.30778 & 13.6058 & 70.0185 & 181.9531 & 0.02554986 & 73.2 & 34.32 & 0.58\\
  J123156+035316 & 97234 & 187.98655 & 3.88782 & 13.6432 & 59.2068 & 166.4686 & 0.01613459 & 62.9 & 33.99 & 0.58\\
  J123209-010543 & 1126897 & 188.04071 & -1.0955 & 14.2598 & 68.0712 & 130.4373 & 0.00917699 & 52.6 & 33.61 & 0.58\\
  J123228+002315 & 4105156 & 188.11709 & 0.38769 & 10.06 & 58.6929 & 170.8812 & 0.00624733 & 12.7 & 30.52 & 0.58\\
  J123236+023943 & 41599 & 188.15304 & 2.66221 & 11.4103 & 74.4606 & 142.1978 & 0.00693114 & 16.7 & 31.11 & 0.58\\
  J123244+000656 & 41618 & 188.18667 & 0.11568 & 7.074 & 83.1277 & 303.1783 & 0.00490181 & 9.5 & 29.89 & 0.58\\
  J123249+031748 & 41625 & 188.20644 & 3.2967 & 12.0076 & 52.43 & 240.9763 & 0.01779495 & 59.7 & 33.88 & 0.58\\
  J123250+054744 & 41627 & 188.21004 & 5.79565 & 12.6288 & 47.6238 & 282.9161 & 0.04053226 & 107.6 & 35.16 & 0.58\\
  J123257+043441 & 41647 & 188.23937 & 4.57811 & 13.5008 & 61.6809 & 80.6525 & 0.00524331 & 14.9 & 30.87 & 0.58\\
  J123320+013119 & 41700 & 188.33727 & 1.522 & 12.8429 & 66.78 & 72.9055 & 0.00672987 & 9.1 & 29.8 & 0.58\\
  J123329+034732 & 41716 & 188.37108 & 3.79235 & 12.5449 & 55.8691 & 103.8952 & 0.00414349 & 15.5 & 30.96 & 0.58\\
  J123407+023905 & 6656564 & 188.53133 & 2.65142 & 6.664 & 77.7896 & 363.2169 & 0.00692832 & 11.1 & 30.22 & 0.58\\
  J123422+021914 & 41816 & 188.59283 & 2.32075 & 10.8866 & 78.8932 & 116.176 & 0.00703584 & 8.9 & 29.76 & 0.58\\
  J123427+021108 & 41823 & 188.61479 & 2.18573 & 7.1732 & 70.1635 & 332.7434 & 0.00715977 & 11.9 & 30.37 & 0.58\\
  J123554-015112 & 41998 & 188.97833 & -1.85339 & 12.6614 & 54.8115 & 264.2974 & 0.02412559 & 96.1 & 34.91 & 0.58\\
  J123555-025405 & 41982 & 188.9812 & -2.90157 & 11.45 & 50.9391 & 315.5279 & 0.0249257 & 76.9 & 34.43 & 0.58\\
  J123609+073152 & 1324268 & 189.03896 & 7.53129 & 11.8096 & 48.4461 & 363.4753 & 0.0251558 & 118.6 & 35.37 & 0.58\\
  J123630+001315 & 1160233 & 189.12733 & 0.22084 & 13.7662 & 52.5855 & 171.2284 & 0.01355351 & 70.2 & 34.23 & 0.58\\
  J123641+030621 & 42080 & 189.17337 & 3.10604 & 13.3826 & 67.8558 & 129.5563 & 0.0059399 & 34.7 & 32.7 & 0.58\\
  J123651-003515 & 1139671 & 189.21624 & -0.58766 & 14.973 & 50.4627 & 70.0198 & 0.00971782 & 22.5 & 31.76 & 0.58\\
  J123703+065533 & 42108 & 189.26346 & 6.92604 & 11.675 & 55.1147 & 101.1826 & 0.00658373 & 9.9 & 29.98 & 0.58\\
  J123734+042150 & 42152 & 189.3924 & 4.36412 & 10.6519 & 56.693 & 204.6089 & 0.01881 & 23.4 & 31.85 & 0.58\\
  J123742-033505 & 42153 & 189.42797 & -3.58497 & 16.4708 & 62.1144 & 75.8018 & 0.02525376 & 52.1 & 33.59 & 0.6\\
  J123748-012038 & 42173 & 189.45018 & -1.34409 & 12.8769 & 51.1762 & 68.0291 & 0.00629811 & 8.1 & 29.55 & 0.58\\
  J123802+001036 & 1159012 & 189.51118 & 0.17674 & 14.8585 & 46.8753 & 139.7513 & 0.01457574 & 79.0 & 34.49 & 0.58\\
  J123805-000137 & 42202 & 189.52254 & -0.027 & 10.6346 & 64.7026 & 343.9878 & 0.01365786 & 62.2 & 33.97 & 0.58\\
  J123805-002419 & 1144183 & 189.52404 & -0.40528 & 14.7913 & 75.3069 & 185.0515 & 0.02660916 & 130.4 & 35.58 & 0.61\\
  J123827+041912 & 42241 & 189.61541 & 4.31999 & 8.4081 & 71.4757 & 255.2234 & 0.00376112 & 12.7 & 30.52 & 0.58\\
  J123835+012413 & 42255 & 189.64817 & 1.40361 & 11.7301 & 76.4072 & 292.1839 & 0.01848818 & 75.6 & 34.39 & 0.58\\
  J123912+060038 & 42319 & 189.80195 & 6.01063 & 10.0503 & 55.6433 & 350.0738 & 0.00921195 & 49.1 & 33.46 & 0.58\\
  J123917-003149 & 42336 & 189.82234 & -0.5305 & 9.3979 & 73.7025 & 201.0799 & 0.00470376 & 12.7 & 30.52 & 0.58\\
  J123922+045608 & 42340 & 189.84246 & 4.93578 & 12.2544 & 48.6094 & 113.3002 & 0.00651829 & 16.0 & 31.02 & 0.58\\
  J123928+041549 & 42354 & 189.8694 & 4.26381 & 12.241 & 70.0185 & 277.7179 & 0.05026681 & 86.9 & 34.7 & 0.58\\
  J123949+014016 & 42393 & 189.95774 & 1.67115 & 11.9926 & 62.0422 & 107.552 & 0.00521402 & 12.9 & 30.55 & 0.58\\
  J124002-010258 & 1128047 & 190.01187 & -1.04943 & 13.3498 & 65.0609 & 77.1982 & 0.00646665 & 12.8 & 30.54 & 0.58\\
  J124002-033623 & 42409 & 190.00896 & -3.60656 & 13.5385 & 59.938 & 114.3869 & 0.00962714 & 29.4 & 32.35 & 0.58\\
  J124109+041725 & 1266160 & 190.28802 & 4.2904 & 14.3298 & 78.463 & 99.0002 & 0.02999321 & 32.2 & 32.54 & 0.59\\
  J124111-013526 & 42538 & 190.29587 & -1.59063 & 12.8232 & 60.9566 & 205.8903 & 0.0144801 & 64.5 & 34.05 & 0.58\\
  J124111+012435 & 42542 & 190.29831 & 1.40986 & 11.6131 & 51.57 & 211.9055 & 0.00680496 & 39.0 & 32.96 & 0.58\\
  J124121+001315 & 1160198 & 190.34052 & 0.22109 & 15.404 & 64.7026 & 141.577 & 0.01714687 & 104.1 & 35.09 & 0.62\\
  J124123-030327 & 42559 & 190.34617 & -3.0577 & 11.5192 & 65.9205 & 98.5783 & 0.00597682 & 8.8 & 29.71 & 0.58\\
  J124128-031517 & 91219 & 190.36976 & -3.25471 & 14.7345 & 72.5815 & 111.0944 & 0.00714245 & 48.3 & 33.42 & 0.58\\
  J124232-000459 & 42689 & 190.63607 & -0.08324 & 8.7895 & 64.4158 & 239.4807 & 0.00687653 & 13.4 & 30.64 & 0.58\\
  J124237-020708 & 3106365 & 190.65649 & -2.11895 & 14.4447 & 48.4461 & 265.9249 & 0.04714775 & 220.9 & 36.72 & 0.58\\
  J124240+012047 & 42709 & 190.66927 & 1.34646 & 11.0747 & 53.5128 & 325.8747 & 0.01788554 & 68.8 & 34.19 & 0.58\\
  J124257-011345 & 42747 & 190.73947 & -1.22941 & 10.3528 & 76.647 & 315.5301 & 0.01184754 & 46.4 & 33.33 & 0.58\\
  J124317-003841 & 42791 & 190.82337 & -0.64492 & 10.2494 & 72.2114 & 303.5107 & 0.00996205 & 41.1 & 33.07 & 0.58\\
  J124333+014827 & 92947 & 190.88767 & 1.8077 & 12.2973 & 61.6086 & 173.919 & 0.04960635 & 36.8 & 32.83 & 0.58\\
  J124350-003344 & 42847 & 190.96079 & -0.56243 & 9.5363 & 48.4461 & 265.9249 & 0.00990722 & 23.0 & 31.81 & 0.58\\
  J124421+042536 & 42892 & 191.08944 & 4.42672 & 12.0215 & 60.6661 & 303.9758 & 0.02972201 & 93.2 & 34.85 & 0.58\\
  J124428-030017 & 42909 & 191.11791 & -3.0048 & 11.1596 & 49.7428 & 453.3832 & 0.02503503 & 133.7 & 35.63 & 0.58\\
  J124428+002815 & 42910 & 191.119 & 0.47091 & 10.9498 & 79.3316 & 103.7941 & 0.00506443 & 7.4 & 29.36 & 0.58\\
  J124433-021916 & 42921 & 191.14097 & -2.32119 & 12.8803 & 55.1904 & 133.9742 & 0.00643985 & 29.3 & 32.34 & 0.58\\
  J124508-002747 & 42975 & 191.28452 & -0.46322 & 6.6948 & 66.1353 & 422.087 & 0.00619415 & 14.9 & 30.87 & 0.58\\
  J124531-003203 & 3110155 & 191.38242 & -0.53434 & 10.5016 & 58.0296 & 157.9589 & 0.00652746 & 13.4 & 30.64 & 0.58\\
  J124548-002600 & 1143451 & 191.45039 & -0.43347 & 13.7782 & 53.6665 & 74.4803 & 0.00670022 & 14.6 & 30.82 & 0.58\\
  J124608+042643 & 1268285 & 191.53441 & 4.4453 & 13.9988 & 65.8488 & 229.0488 & 0.02473573 & 135.6 & 35.66 & 0.58\\
  J124645+055723 & 43106 & 191.69058 & 5.95643 & 10.8463 & 55.9442 & 143.6344 & 0.0038936 & 13.1 & 30.59 & 0.58\\
  J124651-013308 & 3295908 & 191.71291 & -1.55242 & 14.837 & 56.7677 & 69.3402 & 0.02402866 & 20.8 & 31.59 & 0.59\\
  J124701-013444 & 43128 & 191.75426 & -1.57889 & 12.1334 & 49.8231 & 198.9382 & 0.01021996 & 44.0 & 33.22 & 0.58\\
  J124717+002405 & 43153 & 191.82103 & 0.40156 & 11.0785 & 60.6661 & 156.0027 & 0.04758206 & 17.1 & 31.16 & 0.58\\
  J124753-011107 & 43198 & 191.97243 & -1.18529 & 11.8646 & 69.8012 & 349.4939 & 0.02458987 & 113.0 & 35.26 & 0.58\\
  J124800+042609 & 3994854 & 192.0002 & 4.43583 & 12.4899 & 56.2443 & 155.1573 & 0.00450877 & 32.4 & 32.55 & 0.58\\
  J125038+012749 & 43470 & 192.66139 & 1.46365 & 10.874 & 79.7794 & 175.7894 & 0.00535511 & 19.5 & 31.45 & 0.58\\
  J124808+033759 & 5808028 & 192.03338 & 3.63328 & 14.7059 & 54.7356 & 93.0806 & 0.02466785 & 34.1 & 32.67 & 0.7\\
\hline\end{longtable}

\twocolumn

%%%%%%%%%%%%%%%%%%%%%%%%%%%%%%%%%%%%%%%%%%%%%%%%%%

%%%%%%%%%%%%%%%%% APPENDICES %%%%%%%%%%%%%%%%%%%%%

% \appendix

% \section{HI spectra material used for Tully-Fisher distances}

% \textcolor{red}{Khaled here include the HI spectra tables}

%%%%%%%%%%%%%%%%%%%%%%%%%%%%%%%%%%%%%%%%%%%%%%%%%%

% Don't change these lines
\bsp	% typesetting comment
\label{lastpage}
\end{document}